\documentclass[12pt]{article}
\usepackage{jheppub}

\usepackage{bm} 
\usepackage{changes}
\numberwithin{equation}{section}

\hypersetup{
	unicode,	
	colorlinks,
	citecolor=[rgb]{0,.3,.5}, 
	linkcolor=[rgb]{0,.3,.5},
	urlcolor=[rgb]{0,.3,.5}, 
	bookmarksopen=true,
	bookmarksopenlevel=\maxdimen,
}

\def\un#1{\underline #1}

\def\on#1#2{{\buildrel{\mkern2.5mu#1\mkern-2.5mu}\over{#2}}}

\def\dt#1{\smash{\on{\hbox{\bf .}}{#1}}\vphantom{#1}}     

\newcommand{\hyperlabel}[1]{ \hypertarget{#1}{}\label{#1} }

\definecolor{PineGreen}{rgb}{0.0, 0.47, 0.44}
\definecolor{Orange}{rgb}{0.9, 0.4, 0.2}
\definecolor{Red}{rgb}{0.77, 0.01, 0.2}
\definecolor{WillsGrey}{rgb}{0.4, 0.4, 0.5}

\def\physical#1{ {\color{PineGreen} { {#1}} } }
\def\auxiliary#1{ {\color{Orange} {\bm {#1}} } }
\def\FS#1{ { {\physical{\bm {#1}} } } }
\def\compensating#1{ {\color{Red}  #1} }
\def\vanishing#1{ {\color{Red} {\bm {#1}} } }
\def\map#1{ {\color{WillsGrey} {{#1}} } }
\def\combo#1{ {\color{WillsGrey} {#1}} }

\def\gravitonXX{\physical{h} }
\def\gravitonXY{\physical{h} }
\def\gravitonYY{\physical{h} }

\def\frameXX{ \hyperlink{E:frameXX} {\physical{e}} }
\def\frameXY{ \hyperlink{E:frameXY} {\physical{e}} }
\def\frameYX{ \hyperlink{E:frameYX} {\physical{e}} }
\def\frameYY{ \hyperlink{E:frameYY} {\physical{e}} }

\def\spinYXX{ \hyperlink{E:spinYXX} {\physical{\omega}} }

\def\CXXX{ \hyperlink{E:CXXX} {\physical{C}} }
\def\CXXY{ \hyperlink{E:CXXY} {\physical{C}} }
\def\CXYY{ \hyperlink{E:CXYY} {\physical{C}} }
\def\CYYY{ \hyperlink{E:CYYY} {\physical{C}} }

\def\grinoXX{ \hyperlink{E:grinoXX} {\physical{\psi}} }
\def\grinoXY{ \hyperlink{E:grinoXY} {\physical{\psi}} }
\def\grinoYX{ \hyperlink{E:grinoYX} {\physical{\psi}} }
\def\grinoYY{ \hyperlink{E:grinoYY} {\physical{\psi}} }

\def\spinorYY{ \hyperlink{E:spinorYY} {\physical{\chi}} }
\def\spinorYYY{ \hyperlink{E:spinorYYY} {\physical{\chi}} }

\def\d{ \hyperlink{E:d} {\auxiliary{d_X}} }
\def\dX{ \hyperlink{E:dX} {\auxiliary{d}} }
\def\dY{ \hyperlink{E:dY} {\auxiliary{d}} }
\def\dYY{ \hyperlink{E:dYY} {\auxiliary{d}} }

\def\m{ \hyperlink{E:m} {\FS{m}} }
\def\f{ \hyperlink{E:f} {\auxiliary{f}} }

\def\l{ \hyperlink{E:grinEoM} {\auxiliary{\lambda}} }
\def\y{ \hyperlink{E:y} {\auxiliary{y}} }
\def\t{ \hyperlink{E:t} {\auxiliary{t}} }
\def\r{ \hyperlink{E:r} {\auxiliary{\rho}} }

\def\surfeit{ \compensating{\eta} }
\def\compX{ \compensating{G} }
\def\compY{ \compensating{H} }

\def\calibration{ \physical{F} }

\def\KK{ \hyperlink{E:KK}{\combo{\mathcal W}} }
\def\E{ \hyperlink{E:E}{\combo{E}} }
\def\F{ \hyperlink{E:F}{\combo{F}} }
\def\G{ \hyperlink{E:G}{\combo{G}} }
\def\H{ \hyperlink{E:H}{\combo{H}} }

\def\J{ \hyperlink{E:J}{\combo{J}} }
\def\P{ \hyperlink{E:P}{\combo{P}} }
\def\R{ \hyperlink{E:R} { {\FS R} } }
\def\S{ \hyperlink{E:S}{\combo{S}} }
\def\T{ \hyperlink{E:T} {\combo{\widehat{H}}} }
\def\W{ \hyperlink{E:W} {\combo{W}} }
\def\hatW{ \hyperlink{E:hatW} {\combo{\widehat W}} }
\def\X{ \hyperlink{E:X} {\combo{X}} }
\def\Z{ \hyperlink{E:Z}{\combo{\widehat{F}}} }

\def\Ga{ \hyperlink{E:Ga} { \FS{\mathit \Gamma}} } 

\def\Ztorsion{ \hyperlink{E:Ztorsion} {\FS{Z}} }

\def\FSXXXX{\hyperlink{E:FSXXXX} {\FS{G}} }
\def\FSXXXY{\hyperlink{E:FSXXXY} {\FS{G}} }
\def\FSXXYY{\hyperlink{E:FSXXYY} {\FS{G}} }
\def\FSXYYY{\hyperlink{E:FSXYYY} {\FS{G}} }
\def\FSYYYY{\hyperlink{E:FSYYYY} {\FS{G}} }
\def\Curvature{ {\FS{R}} }

\def\Torsion{ {\FS{T}} }

\def\Curvature{ {\FS{R}} }

\def\EoMgrino{ \hyperlink{E:EoMgrino} {\vanishing{ E_{\Psi}}} }
\def\EoMX{ \hyperlink{E:EoMX} {\vanishing{ {E_{U}}}} }
\def\EoM{ \hyperlink{E:EoM} {\vanishing{ E_{X}}} }
\def\EoMY{ \hyperlink{E:EoMY} {\vanishing{ E_{\mathcal V}}} }
\def\EoMYY{ \hyperlink{E:EoMYY} {\vanishing{ E_{V}}} }
\def\EoMYYY{ \hyperlink{E:EoMYYY} {\vanishing{ E_{\Phi}}} }

\def\hmap{ \hyperlink{E:hmap} {\map{h}} }
\def\tmap{ \hyperlink{E:tmap} {\map{t}} }
\def\hitchin{ \hyperlink{E:HitchinMetric} {\map{G}} }

\preprint{MI-TH-211}

\title{
Components of Eleven-dimensional Supergravity with Four Off-shell Supersymmetries
} 

\author[a]{Katrin Becker,}
\author[a]{Daniel Butter,}
\author[a]{William D. Linch III,}
\author[a]{and Anindya Sengupta.}
\affiliation[a]{
George P. and Cynthia Woods
Mitchell Institute for 
Fundamental Physics and Astronomy, \\
Texas A\&{}M University.\\
College Station, TX 77843, USA
} 

\emailAdd{kbecker@physics.tamu.edu}
\emailAdd{dbutter@tamu.edu}
\emailAdd{wdlinch3@gmail.com}
\emailAdd{anindya.sengupta@tamu.edu}

\abstract{
We derive the component structure of 11D, $N=1/8$ supergravity linearized around eleven-dimensional Minkowski space. This theory represents 4 local supersymmetries closing onto 4 of the 11 spacetime translations without the use of equations of motion. It may be interpreted as adding $201$ auxiliary bosons and $56$ auxiliary fermions to the physical supergravity multiplet for a total of $376+376$ components. These components and their transformations are organized into representations of $SL(2;\mathbf C)\times G_2$.
}

\begin{document}
\maketitle

\newpage
\section{Introduction}
\label{S:Intro}

Supersymmetric field theories are invariant under a set of transformations taking bosons to fermions and {\it vice versa}. By definition, these supersymmetry transformations form a Lie superalgebra extending the algebra of Poincar\'e transformations to a super-Poincar\'e algebra. By construction, the fields in the theory are linear representations of the bosonic Poincar\'e subalgebra. Ideally, this {supermultiplet} of fields furnishes a (possibly reducible) linear representation of the full supersymmetry algebra. For this to be true, it is necessary for the commutator of two supersymmetries|as realized on a component field|to close onto the translations on that field (up to other bosonic symmetries and possibly gauge transformations).

\paragraph{On- and Off-shell Supersymmetry}
However, it is generally the case that this requirement fails and---strictly speaking---the fields do not furnish a linear representation of the supersymmetry algebra.
If the action is invariant under the fermionic transformations that extend the Poincar\'e symmetries to a consistent superalgebra, how can the fields fail to be a representation? An important observation in this context is that the failure of the supersymmetries to close on the fields is by terms that vanish if the equations of motion are imposed.\footnote{Such terms are called {\em trivial symmetries}. In a canonical treatment of symmetries, in which there is a Poisson bracket on the space of fields, these symmetries are canonical transformations generated by the action $S$ itself. They are trivial, in the sense that $\delta S \sim \{ S, S\}_{\mathrm {PB}} \equiv 0$.} 
In this context, the supersymmetry is said to {\em close on-shell} only, and the set is referred to as an {\em on-shell  supermultiplet}.
{\em This should not be interpreted to mean that the component fields in the theory are required to be on-shell}; only that the supersymmetries do not close on the off-shell realization. 

In the ideal case, this complication does not arise: Successive supersymmetry transformations close properly on the fields irrespective of any field equations, and the collection of fields furnishes a linear representation. When this fortuitous situation is realized, it is emphasized by referring to it as {\em off-shell supersymmetry}. When we compare these off-shell theories to the on-shell ones described previously, we notice that they avoid the obstruction to closure with extra fields: Roughly, the terms in the transformations of components that give rise to obstructions are replaced with new fields. 

\paragraph{Auxiliary Fields and the Off-shell Problem}
Under supersymmetry transformations, the new component fields transform into the obstruction. 
Such fields are called {\em auxiliary fields}. 
In two-derivative actions, they have algebraic equations of motion (no derivatives on them), so they do not contribute dynamical degrees of freedom. In principle, they can be integrated out of the path integral (or set to their classical values), but removing them changes the supersymmetry transformations which now close only on-shell.

The problem of extending on-shell formulations of supersymmetric theories to off-shell supersymmetry by adding auxiliary fields is known as {\em the off-shell problem}. It is considered important for various reasons ranging from the purely mathematical representation theory of supersymmetry algebras to phenomenological supersymmetric model building. (In the case of most interest to us in this paper, we intend to use off-shell supersymmetry to study self-interactions in the form of higher-derivative corrections to the M-theory effective action.)

A considerable amount of effort has been devoted to solving the off-shell problem. An immediate indication of the inadequacy of any step-wise approach attempting to add suitable component fields and transformations was provided by Siegel and Ro\v{c}ek \cite{Siegel:1981dx} (see also sec. 2.4.3 of \cite{Galperin:2001uw}): For {\em any non-real representation} in theories with eight or more supersymmetries ({\it e.g.}\ hypermuliplets), any potential solution to this problem requires an infinite number of fields. The proof is essentially a counting argument comparing the dimensions of off-shell Fock representations of the supertranslation algebra to those of superfield representations.

\paragraph{Manifest Supersymmetry}
Just as the translation part of the Poincar\'e algebra is represented on fields by coordinate derivatives, so too can the supersymmetries be geometrized to translations in fermionic directions. Introducing coordinates for these directions, we can consider {\em superfields} as functions of both the even and odd coordinates. The component fields of a supermultiplet then appear as coefficients in the Taylor expansion of the superfields in the odd variables. 

By construction, superfields are field representations of the super-Poincar\'e algebra (just as the component fields are so for the Poincar\'e subalgebra), and the supersymmetry is said to be {\em manifest}. However, these are generally (highly-)reducible representations, so we must reduce them in a way that is compatible with manifest supersymmetry. 
To this end, we can impose combinations of reality conditions, gauge symmetry, and multiplet-reducing constraints compatible with superspace. In the latter, we use the supertranslations generated by fermionic covariant derivatives $D$ to impose covariant constraints such as $D\Phi = 0$, $D^2 \Phi =0$, etc. These are often called {\em shortening conditions}, because they imply that $\Phi$ does not depend on some of the odd coordinates thereby shortening its Taylor expansion.

The problem of finding irreducible representations of the super-Poincar\'e algebra thus becomes one of understanding the possible shortening conditions. This is non-trivial because the superalgebra implies that successive $D$s can generate bosonic derivatives $\{D, D\}\sim -2i \partial$, so imprudently-chosen constraints generate kinetic operators generalizing those of d'Alembert and Dirac. In this event, the multiplet is on-shell again in the sense described above.
But note that {\em in superspace it further implies the superfield equations of motion directly on the representation}. In other words, we cannot take such a manifestly supersymmetric representation off-shell in any sense; the construction of an action principle using such a representation is moot.

Much work in the superspace literature has been devoted to finding clever shortening conditions by enlarging the superspace with appropriate bosonic spaces. These naturally circumvent the counting arguments by introducing field dependence on bosonic coordinates geometrizing the R-symmetry transformations \cite{Galperin:1984av, Lindstrom:1989ne,Galperin:2001uw} or on bosonic ghost-like variables ({\it e.g.}\ Lorentz harmonics \cite{Galperin:1991gk, Delduc:1991ir} or pure spinor \cite{Berkovits:2000fe,Cederwall:2010tn,Berkovits:2018gbq}). This approach can be generalized and systemized using supercosets \cite{Siegel:1999ew}.

Alternatively, we can attempt to work around the problem using a superspace with a {\em smaller structure group} than the maximal one. Examples of this include light-cone superspace \cite{Mandelstam:1982cb} and its generalizations ({\it e.g.}\ \cite{Marcus:1983wb, Berkovits:1993hx}). In these reductions, new shortening conditions can be defined that are not manifestly covariant under the original symmetry group, but also do not imply any dynamical equations. Such off-shell representations may then be combined into a supermultiplet representing the larger bosonic group linearly, albeit not manifestly.

\paragraph{11D, $N=1/8$ Superspace}
An attempt to classify the possible structure group reductions appropriate to 16 supercharges was made in \cite{Evans:1994np}.
In previous work, we applied a similar approach to eleven-dimensional supergravity based on the reduction $Spin(10,1)\to SL(2, \mathbf C) \times G_2$ \cite{Becker:2018phr, Becker:2017zwe, Becker:2017njd, Becker:2016edk, Becker:2016rku, Becker:2016xgv}. 
Besides being the most familiar and phenomenologically relevant, this superspace has the significant advantage of requiring only finitely-many auxiliary fields (and has the largest super-Poincar\'e symmetry with this property \cite{Siegel:1981dx}).
In this formulation the component fields of eleven-dimensional supergravity are embedded into a set of so-called {\em prepotential superfields} satisfying no constraints other than the reality and shortening conditions ({\it i.e.} chiral superfields) allowed off-shell. 
Besides the 11D frame $e_{\bm m}{}^{\bm a}$, 3-form $C_{\bm {mnp}}$, and gravitino $\psi_{\bm m}{}^{\bm \alpha}$, these superfields contribute a finite set of auxiliary fields guaranteeing that 4 of the 32 supersymmetries close off-shell on the set of physical + auxiliary fields.

In this paper, we will construct superfields whose $\theta$-independent components are precisely the physical 11D gauge connections ({\it i.e.}\ the frame, 3-form, and gravitino). 
The manifest supersymmetry transformations of the component fields are effected by acting with the $SL(2, \mathbf C)$ superspace derivatives $D$ on these superfields. 
Besides mixing the bosonic gauge fields, this action generates the auxiliary fields in a form familiar from the standard component ``tensor calculus''. 
We use this to explicitly identify the spectrum of auxiliary fields (or, more precisely, the superfields that have the auxiliary components as their $\theta$-independent term).
The supergeometrical tensors of the eleven-dimensional theory ({\it i.e.}\ 4-form field strength, torsion, and curvature 2-form) may then be defined in the na\"ive way by taking the curls of the gauge connections. 
By construction, this spectrum of superfields forms a(n off-shell) representation of the reduced structure supergroup. 

As a result of this analysis, we {\em derive} the set of torsion constraints for the supergeometry with this reduced structure group. 
As it is traditionally done, this set is the starting point for solving the curved superspace Bianchi identities \cite{Gates:1983nr,Wess:1992cp,Buchbinder:1998qv}.
\footnote{It is known that imposing just the dimension zero torsion condition 
$T_{\bm\alpha \bm\beta}{}^{\bm c} \propto (\Gamma^{\bm c})_{\bm \alpha \bm \beta}$
is sufficient to put 11D superspace on-shell \cite{Howe:1997he}. This observation allows one to postulate an ``off-shell'' formulation of 11D supergravity (or at least parametrize higher derivative corrections) by relaxing this constraint and re-solving the Bianchi identities \cite{Cederwall:2000ye}. However, there is some disagreement about this point
\cite{Nishino:1996tw,Gates:2001hf,Gates:2001zz}.
Our result does not speak directly to this issue, because we attempt
to keep only 4 of the 32 supersymmetries off-shell.}
At the end of this lengthy and technical process (which we do not carry out in this paper), one finally arrives at the solution to the constraints which are the off-shell prepotentials. 
Amusingly, then, in the situation we described above and will follow in this paper, we have essentially started with the solution to a set of constraints we do not yet know, and can then {\em derive} these constraints by acting repeatedly with the superspace derivatives on the unconstrained fields. 
Since the prepotentials with which we start are unconstrained, this process is guaranteed to terminate without ever imposing a dynamical constraint. 
Along the way, all the physical potentials, field strengths, and Bianchi identities are derived.

\subsection{Results}
\label{S:Results}

The component fields of eleven-dimensional supergravity consist of a frame $e_{\bm m}{}^{\bm a}$, a gauge 3-form $C_{\bm {mnp}}$, and a gravitino $\psi_{\bm m}^{\bm \alpha}$.
Under linearized supersymmetry, they transform into each other as
\begin{align}
\label{E:LinSUSY}
\delta_{\bm \epsilon} e_{\bm a }{}^{\bm b} &= \bar {\bm \epsilon} \Gamma^{\bm b} \psi_{\bm a}
\cr
\delta_{\bm \epsilon} C_{\bm {abc}}  &= 3 \bar {\bm \epsilon} \Gamma_{[\bm {ab} }\psi_{\bm c]}
\cr
\delta_{\bm \epsilon} \psi_{\bm a}{}^{\bm \beta}  &= -\tfrac12 \omega_{\bm a}{}^{\bm{bc}}
	(\Gamma_{\bm {bc}} \bm \epsilon)^{\bm \beta}
	-  \tfrac1{6\cdot 4!} \left( 3 \Gamma^{\bm {bcde}}\Gamma_{\bm a }\bm \epsilon 
		- \Gamma_{\bm a } \Gamma^{\bm {bcde}}\bm \epsilon \right) \bm G_{\bm {bcde}}
~,
\end{align}
where $\bm G := d C$ is the 4-form field strength, and we have flattened all the indices with the background frame $\delta_{\bm m}^{\bm a}$.
As a realization of supersymmetry, these transformations can only close on-shell: 
The degrees of freedom needed for the off-shell counting to match are missing (cf.\ table \ref{T:DoFCounting}).
\begin{table}[t]
{\footnotesize
\begin{align*}
{\renewcommand{\arraystretch}{1.7} 
\begin{array}{|c|c|c|c|c|c|c|c|}
\hline
\textrm{gauge field} 
	& ~~~~\textrm{on-shell}~~~~ 
	& \textrm{off-shell}\\
\hline
\textrm{frame} 
	& ~~\tfrac12 ({\textrm{D}}-1)({\textrm{D}}-2) -1 ~~
	& \tfrac12 {\textrm{D}}({\textrm{D}}-1) \\
	&44  
	& 55 \\
\hline
\textrm{gravitino}
	& 2^{\lfloor{\tfrac {\textrm{D}}2}\rfloor -1} \cdot ({\textrm{D}}-3)
	& ~~2^{\lfloor{\tfrac {\textrm{D}}2}\rfloor} \cdot ({\textrm{D}}-1)~~ \\
	& 128 
	& 320 \\
\hline
p\textrm{-form} 
	& {{\textrm{D}}-2\choose p} 
	& {{\textrm{D}}-1\choose p}  \\
p=3 
	&84 
	& 120 \\
\hline
\hline
~~\textrm{total (D=11)}~~
	& {\color{PineGreen}128+128 }
	&  {\color{Red}175+ 320} \\
\hline	
\end{array}
}
\end{align*}
\caption{\scriptsize Counting of degrees of freedom in D dimensions relevant to Poincar\'e supergravity. (For conformal supergravity, some (gamma-)traces should be subtracted.)
}
\label{T:DoFCounting}
} 
\end{table}
In this paper, we will explain how to fix this mismatch and give explicitly the modifications to these supersymmetry transformations (cf.\ eqs.\ \ref{E:SupersymmetryTransformationsGraviton}, \ref{E:SupersymmetryTransformationsThreeform}, and \ref{E:SupersymmetryTransformationsGravitino}).

To get a representation on which 4 of the 32 supersymmetries close off-shell, we decompose the 11D component fields under
\begin{align}
\label{E:StructureGroupReduction}
Spin(10,1) 
	\to Spin(3,1) \times Spin(7)
	\to Spin(3,1) \times G_2 
	= SL(2, \mathbf C)\times G_2
~.	
\end{align}
For notational convenience, we will label spacetime, polarizations, indices, {\it et cetera} by $M$ for 11D, $X$ for the 4D part, and $Y$ for the 7D part. 
Under the decomposition of the tangent space, 
\begin{align}
\begin{array}{cccclcc}
	e_{\bm m}{}^{\bm a} &\to &
		\frameXX_{m}{}^{a} ~,~ 
		\frameYX_{i}{}^{a} ~,~ 
		\frameXY_{m}{}^{j} ~,~ 
		\frameYY_i{}^j 
&\textrm{with }& a, m = 0,1,2,3,
\\
	\psi_{\bm m}^{\bm \alpha}&\to &
		\grinoXX_{m}^{~\alpha} ~,~ 
		\grinoXY_{m}^{~\alpha i} ~,~ 
		\grinoYX_{j}^{~\alpha } ~,~ 
		\grinoYY_{j}^{~\alpha i} ~,~
&\textrm{and }& \alpha = 1,2,	
\\ 
	C_{\bm {mnp}} &\to &
		\CXXX_{mnp}~,~ 
		\CXXY_{mn \, i} ~,~ 
		\CXYY_{m \, ij}~,~ 
		\CYYY_{ijk}
&\textrm{and }& i,j = 1,\dots, 7	
~.
\end{array}
\end{align}
To avoid introducing too much notation, the $i, j$, \dots indices will be doing quadruple duty as 7D coordinate indices, 7D tangent space indices, $G_2$ indices, and as a label for 7 additional gravitino fields. (We collect the index definitions in table \ref{T:Legend}).
\begin{table}[ht]
{\footnotesize
\begin{align*}
{\renewcommand{\arraystretch}{1.5} 
\begin{array}{|c|c|c|c|c|c|c|c|}
\hline
\textrm{index} & ~~~~\textrm{range}~~~~ & \textrm{description} &~~~\textrm{label}~~~\\
\hline
\bm m, \bm n, \dots & 0, \cdots , 10 & \textrm{11D coordinate} &\\
\bm a, \bm b, \dots & 0, \cdots , 10 & \textrm{11D tangent} &M\\
\bm \alpha, \bm \beta, \dots & 1, \cdots , 32 & \textrm{11D spinor} &\\
\hline
m, n, \dots & 0, 1,2 , 3 & \textrm{4D coordinate} &\\
a, b, \dots & 0, 1,2 , 3 & \textrm{4D tangent} &X \\
~~\alpha, \beta , \dots , \dt \alpha, \dt \beta \dots & 1, 2 & \textrm{4D spinor} &\\
\hline
i,j,\dots & 1,\cdots , 7 & ~~\textrm{7-component label} ~~&Y\\
\hline
\end{array}
}
\end{align*}
\caption{\scriptsize Legend of indices used in this work. To avoid having even more indices, those in the various $\bm7$-dimensional representations ($GL(7)$ coordinate, $SO(7)$ tangent, $G_2$ representation, and label for seven gravitini), have all been identified. 
}
\label{T:Legend}
} 
\end{table}
Using the $G_2$ structure to trade the 7D graviton polarizations $\gravitonYY_{ij}$ for a stable 3-form $F_{ijk}$, these components embed into a set of superfields called {\em prepotentials} \cite{Becker:2017zwe}, as summarized in table \ref{T:Components}.

The prepotential superfields are unconstrained superfields or constrained only to be real or chiral.\footnote{ Such constraints are innocuous: Besides factors of $\bar D^2$ arising from the variation of chiral fields, these fields are unconstrained as integration variables in a path integral. (For this reason, we will sometimes sloppily refer to them as unconstrained.)}
The components that arise in the Taylor expansion of the fields are listed under the superfield. They have been separated into three groups:
\begin{itemize} 
	\item {\color{PineGreen} Gauge fields} transform canonically. Their (double) curls are field strength, torsion, and curvature 2-form components. Using table \ref{T:DoFCounting}, they add up to $128|128$ components on-shell, but off-shell (modulo gauge transformation), we find a mismatch between $175$ bosonic and $320$ fermionic components.
	\item {\color{Orange} Auxiliary fields} are gauge invariant, but do not propagate. Due to this, it is easily verified that they contribute $201$ bosonic and $56$ fermionic degrees of freedom all of which are off-shell.
	\item {\color{Red} Compensators} suffer St\"uckelberg shifts under ``pregauge'' transformations. They do not appear in the spectrum of gauge(-invariant) superfields, so they do not contribute any degrees of freedom at all.
\end{itemize}

The compensating fields are artifacts of the reduction of the structure supergroup implied by \eqref{E:StructureGroupReduction}. 
This reduction is crucial to the representation of the off-shell supersymmetries, but it has no geometrical eleven-dimensional meaning. (Presumably it has some pregeometrical meaning, but the off-shell pregeometry for 11D is precisely the thing we do not know.)
This situation is precisely analogous to 4D, $N=1$ Poincar\'e supergravity itself, which is described in terms of conformal supergravity (which is not a physical symmetry) coupled to a scalar superfield called the {\em conformal compensator} and denoted $\G$ \eqref{E:G} in this paper.

Relatedly, the auxiliary fields that appear in the Taylor expansion of the prepotentials are---strictly speaking---not invariant under all of the pregeometrical symmetries. 
However, the parts of the pregeometrical transformations under which they are not covariant are again artifacts of the structure group reduction having no eleven-dimensional interpretation. 
Therefore, it is possible to covariantize these auxiliary components by mixing in some parts of the other prepotentials. 
This is done explicitly in section \ref{S:11DSuperfields}, but the details are not important here.

\begin{table}[t]
{\footnotesize
\begin{align*}
~\hspace{-13mm}
{\renewcommand{\arraystretch}{1.5} 
\begin{array}{|c|c|c|c|c|c|c|c|c|}
\hline
& U^a & \Psi_i^\alpha & \mathcal V^i  & X & \Sigma_{\alpha i} & V_{ij} &  \Phi_{ijk} &~~\textrm{sdim}_{\mathbf R} ~~\\
~\textrm{prepotential}~& \textrm{real} & \textrm{spinor} & ~~\textrm{real}~~ &  ~~\textrm{real}~~ &  ~~\textrm{chiral}~~ &  \textrm{real} &  \textrm{chiral}&\\
\hline
\physical{\textrm{physical }}
	& ~~\gravitonXX_{ab} , \grinoXX_a{}^\beta ~~
	& \grinoXY_a{}^{\beta j} & \gravitonXY_a^i & \CXXX_{abc} 
	& \CXXY_{ab\,i} & ~~\CXYY_{a\, ij}, \spinorYY_{\alpha\, ij} ~~
	& ~~\CYYY_{ijk}, \gravitonYY_{ij}, \spinorYYY_{\alpha\, ijk}~~&\\
\auxiliary{\textrm{auxiliary }} & \dX^a 
	&~~ \l^i_\alpha, \y_i^a, \t_i{}^{\alpha \beta} ,\r_i^\alpha ~~
	& \dY^i & \d &  & \dYY_{ij} & \f_{ijk} &\\
~~\compensating{\textrm{compensating}}~~ 
	&&&\surfeit^i_\alpha &\compX,\surfeit'_\alpha 
	&\compY_i,\surfeit''_{\alpha i}&&&
	\\
\hline	
\textrm{on-shell} & 2|2 & 0|14 & 14|0 & 0|0 & 7|0 & 42|42 & 63|70 & 128|128 \\
\textrm{off-shell} & (\physical{6}+\auxiliary{4})|\physical{12} & \auxiliary{98}|(\physical{84}+\auxiliary{56}) & (\physical{21}+\auxiliary{7})|0 & (\physical{1}+\auxiliary{1})|0 & \physical{21}|0 & (\physical{63}+\auxiliary{21})|\physical{84} & (\physical{63}+\auxiliary{70})|\physical{140} &   376|376\\
\hline
\end{array}
}
\end{align*}
\caption{\scriptsize The unconstrained ``prepotential'' superfields of the 11D, $N=1/8$ supergravity multiplet and their physical, auxiliary, and compensating component content. The components of the gravitino of the type $\psi_i{}^\beta$ and $\psi_i{}^{\beta j}$ are encoded in $\chi_{\alpha ij}$ and $\chi_{\alpha ijk}$ \eqref{E:spinor}. The real superdimension contributed by each superfield is computed modulo gauge transformation. The superdimension of the auxiliary field space adds up to $201|56$. 
}
\label{T:Components}
} 
\end{table}

\paragraph{Supersymmety Transfomations}
Once the eleven-dimensional superfields corresponding to the components in table \ref{T:Components} have been defined, we are in a position to compute their supersymmetry transformations.
In superspace, supersymmetry transformations of component fields result from infinitesimal translations in the odd directions---in other words---from acting with the fermionic covariant derivatives. 
The result of this calculation is that the graviton transformation is unchanged from the on-shell rule:
\begin{align}
\label{E:SupersymmetryTransformationsGraviton}
\delta_\epsilon \gravitonXX_{ab} &= 
	2i \epsilon \sigma_{(a} \bar \grinoXX_{b)}
	+\mathrm{h.c.}
\cr
\delta_\epsilon \gravitonXY_{a i} &= 
	- i \epsilon\grinoXX_{a\,i} +i \epsilon \sigma_{a} \bar \grinoYX_{i} 
	+\mathrm{h.c.}
\cr
\delta_\epsilon \gravitonYY_{ij} &= 
	-2i \epsilon\grinoYY_{(i, j)}
	+\mathrm{h.c.}
\end{align}
($SL(2, \mathbf C)$ spinor index contractions are implied.)
The supersymmetry transformation of the 3-form acquires a correction by the dimension-1/2 auxiliary field $\l^i_\alpha$ of the gravitino multiplet (cf.\ table \ref{T:Components}, or eq.\ \ref{E:grinEoM} for the explicit expression):\footnote{Some of the gravitino components can be shifted by the gauge-invariant dimension-1/2 auxiliary field  $\l^i_\alpha$, and this type of correction could also have appeared in the graviton transformation. Because this field vanishes on-shell, this leads to an ambiguity in the definition of the off-shell gravitini. The definition of the gravitino we make below (see eq.\ \ref{E:grino}) ensures that $\l^i_\alpha$ appears in only one place in the \emph{off-shell} transformations \eqref{E:SupersymmetryTransformationsGraviton} and
\eqref{E:SupersymmetryTransformationsThreeform}.
} 
\begin{align}
\label{E:SupersymmetryTransformationsThreeform}
\delta_\epsilon \CXXX_{abc} &= -6 \epsilon \sigma_{[ab} \grinoXX_{c]}
	+\mathrm{h.c.}
\cr
\delta_\epsilon \CXXY_{ab\, i} &= - 4i \epsilon \sigma_{[a} \bar \grinoXY_{b] i} 
	-2 \epsilon \sigma_{ab} \grinoYX_{i}
	+\mathrm{h.c.}
\cr
\delta_\epsilon \CXYY_{a\, ij} &= \varphi_{ijk} \epsilon \grinoXY_{a}{}^k 
	-4i\delta_{k[i} \epsilon \sigma_{a} \bar \grinoYY_{j]}{}^k
	-\tfrac16 \varphi_{ijk} \epsilon \sigma_{a} \bar \l_{}^k
	+\mathrm{h.c.}
\cr
\delta_\epsilon \CYYY_{ijk} &= \varphi_{l[ij} \epsilon \grinoYY_{k]}{}^l
	+\mathrm{h.c.}
\end{align}
In these transformations, we define $\delta_\epsilon$ to include a Wess-Zumino gauge transformation|that is, a field-dependent gauge transformation that involves unphysical terms in the superfield multiplet.

Finally, we present the gravitino supersymmetry transformations. 
These allow us to identify the dimension-1 auxiliary fields needed to close the supersymmetry algebra.
For legibility, we suppress the ubiquitous spin connection terms in addition to including compensating Wess-Zumino gauge transformations.
(The complete expressions are presented in appendix \ref{S:GrinoDescendants}.)
\begin{align}
\label{E:SupersymmetryTransformationsGravitino}
\delta_\epsilon \grinoXX_a^\beta &\simeq
	- \tfrac i2  (\epsilon\eta_{ab} + \epsilon \sigma_{ab})^\beta \dX^b
	-\tfrac i2 (\bar \epsilon \bar \sigma_a )^{\beta} \R
\\
\delta_\epsilon \grinoYX_i^\beta &\simeq
	-\tfrac i{12} (\epsilon \sigma^{ab})^\beta \t_{ab\, i}
	-\tfrac i2 \epsilon^\beta \dY_i  
	- \tfrac i{12} (\bar \epsilon \bar \sigma_a )^\beta \Ga_i^{a}
\cr
\delta_\epsilon \grinoXY_{a}{}^{\beta j} 
	&\simeq 
	-\tfrac{5i}{12} (\epsilon\eta_{ab} + \tfrac 85 \epsilon \sigma_{ab} )^\beta\,  \bar \Ga^{bj}
	-i (\bar \epsilon\bar \sigma_a )^\beta \dY^j - \tfrac i6 
	\big[  4 \eta_{ab} \bar \epsilon \bar \sigma_c + i \varepsilon_{abcd} \bar \epsilon \bar \sigma^d 
		\big]^\beta \t^{jbc}
\cr&
-\tfrac14 \epsilon^\beta \big[
		i \Ga^j_a
		-\y^j_a - \bar \y^j_a
	\big]
\cr
\delta_\epsilon \grinoYY_{i}{}^{\beta j} &\simeq
	-i(\epsilon \sigma^{ab})^\beta \pi_{\bm 14} \FSXXYY_{ab\,i}{}^j
	-i (\bar \epsilon \bar \sigma^a)^\beta \left[
		\hmap_i^j(\FSXYYY_{a\, klm})
		+\tfrac i{6}\varphi_i{}^{jk} ( 
			i  \Ga_{a k}
			-\y_{a k}
			)
		\right]
	- \tfrac 12 \epsilon^\beta \Ztorsion_i{}^j
\nonumber	
\end{align}
In these expressions, only certain $G_2$ representations of the 4-form field strength $\FS G_{\bm{abcd}}$ and dimension-1 auxiliary fields appear explicitly. 
For the former, these are the $\bf{14}$-dimensional part of the $[2,2]$-form component $\FSXXYY_{ab\, ij}$, and the $\bf{27}+\bf{1}$ parts of the $[1,3]$-form component $\FSXYYY_{a\, ijk}$. 
(The explicit definitions of these projectors are given in appendix \ref{S:G2}; cf.\ eqs.\ \ref{E:21to14} and \ref{E:hmap}.)
For the latter, the invariants $\R$, $\Ga_i^a$, and $\Ztorsion_i{}^j$ appearing here are combinations of the more basic auxiliary fields $\d$, $\y_i^a$, $\dYY_{ij}$, $\f_{ijk}$ fields in table \ref{T:Components} and components of the 4-form field strength. 
These relations are given explicitly in section \ref{S:Auxiliary} (cf.\ eqs.\ \ref{E:R}, \ref{E:Ga}, and \ref{E:Ztorsion}).

\paragraph{Outline}
This concludes our description of the $376|376$-component multiplet of eleven-dimensional supergravity furnishing a linear representation of the superalgebra with four real supersymmetries. 
In the remainder of the paper, we will explain the derivation of this result.
In the process, we will provide explicit expressions for the gauge superfields, their field strengths, and relations between these under the action by the fermionic superspace derivatives. 

The results are organized as follows:
In the next section, we give the steps starting from the basic prepotentials and arriving at the connection and auxiliary superfields.
In section \ref{S:Dynamics}, we describe the dynamical equations that result from the off-shell action derived in \cite{Becker:2017zwe}. 
These equations relate auxiliary fields to physical components, thereby demonstrating the relation between on- and off-shell field content.
In section \ref{S:Supergeometry}, we consolidate the results for the 4-form, torsion, and curvature field strengths in terms of supergeometry, and conclude the body of the paper. 

What follows are four appendices with conventions beginning with basic results for 4D, $N=1$ superspace (app.\ \ref{S:Superspace}) and 
the essentials of $G_2$ geometry (app.\ \ref{S:G2}). 
This is followed by an appendix \ref{S:Pregauge} of intermediate calculations combining the prepotentials of section \ref{S:Prepotentials} into building blocks with simplified transformations that are used to define all the other fields throughout this work. 
Finally, we present the result of acting with fermionic superspace derivatives in all possible ways on the gravitino components in appendix \ref{S:GrinoDescendants}. 
This demonstrates the consistency of the scheme and generates the supersymmetry transformations of the eleven-dimensional gravitino components reported above in \eqref{E:SupersymmetryTransformationsGravitino}.

\section{Prepotentials}
\label{S:Prepotentials} 

We now begin the task of explicitly identifying all the elements described in the previous sections. 
As implied in table \ref{T:Components}, we embed the components into prepotential superfields
\begin{align} 
\label{E:Prepotentials}
U^a &\sim \cdots 
	+ (\theta \sigma^m \bar \theta) \frameXX_m{}^a 
	+ \theta^2 (\sigma^m \bar \theta)_\alpha \grinoXX_m{}^\alpha 
	+ \bar \theta^2 (\theta \sigma^m )_{\dt \alpha} \bar \grinoXX_m{}^{\dt \alpha }
	+ \theta^2 \bar \theta^2 \dX^a
\cr
\Psi^{\alpha i} &\sim \cdots 
	+ (\theta \sigma^m \bar \theta) \grinoXY_m{}^{\alpha i}
	+ \theta^2 (\bar \theta \bar \sigma_a)^\alpha \y_i^a
	+ \bar \theta^2 (\theta \sigma^{ab})^\alpha \t_{ab i} 
	+ \theta^2 \bar \theta^2 \r^i_\alpha
\cr
\mathcal V^i &\sim \cdots 
	+ (\theta \sigma^m \bar \theta) \frameXY_m^i 
	+ \bar \theta^2 \theta^\alpha \surfeit_\alpha^i
	+ \theta^2 \bar \theta_{\dt \alpha}\bar \surfeit^{\dt \alpha i}
	+ \theta^2\bar \theta^2 \dY^i
\cr
X &\sim \cdots 
	+ \theta^2 \bar \compX
	+ \bar \theta^2 \compX
	+ \theta \sigma_m \bar \theta \mathcal \, \varepsilon^{mnpq} \CXXX_{npq}
	+ \bar \theta^2 \theta^\alpha \surfeit_\alpha
	+ \theta^2 \bar \theta_{\dt \alpha}\bar \surfeit^{\dt \alpha}
	+ \theta^2\bar \theta^2 \d
\cr
\Sigma_i^\alpha &\sim \cdots	
	+ [\theta^\alpha \compY_i + (\theta \sigma^{mn})^\alpha \CXXY_{mn \, i} ] 
	+  \theta^2 \tilde \surfeit_i^\alpha
\cr
V_{ij} &\sim \cdots 
	+  (\theta \sigma^m \bar \theta)  \CXYY_{m\, ij}
	+ \bar \theta^2 \theta^\alpha \spinorYY_{\alpha \, ij} 
	+ \theta^2 \bar \theta_{\dt \alpha}\bar \spinorYY_{ij}^{\dt \alpha}
	+ \theta^2\bar \theta^2 \dYY_{ij}
\cr
\Phi_{ijk} &\sim \CYYY_{ijk} + i \calibration_{ijk}  
	+ \theta^\alpha \spinorYYY_{\alpha\, ijk}
	+ \theta^2 \f_{ijk}
\end{align}
We recall that, in addition to the {\color{PineGreen} gauge fields} transforming canonically, there are {\color{Orange} auxiliary fields} that do not propagate, and {\color{Red} compensating components} that shift under pregauge transformations so they do not appear in the spectrum. 

Our main result is a kind of inversion of this embedding: We will present explicit expressions for 11-dimensional superfields containing these gauge fields and auxiliary fields as the leading term in their Taylor expansions. 
(To avoid introducing yet more notation, we will use the same symbols to denote the component field and the superfield having that component as the leading term in its $\theta$-expansion. 
This should cause no confusion, as henceforth we will always mean the superfield unless explicitly stated otherwise.)
There are no analogous superfields for the compensating components, precisely because these have no lift to 11D.

These superfields are complicated but explicit combinations of the prepotentials that transform only under the physical part of the pregauge transformations (by which we mean those corresponding to 11D diffeomorphisms, supersymmetry, Lorentz transformations, and 3-form gauge transformations). Concretely, the gauge fields must transform as 11D supergravity gauge fields, whereas the auxiliary fields must be invariant.

The procedure to construct the eleven-dimensional superfields is the following:
\begin{enumerate}
	\item We write the linearized transformations of the prepotentials \eqref{E:Prepotentials}. These are collected in appendix \ref{S:Pregauge}. They contain abelian transformations of the M-theory 3-form, a non-abelian gauging thereof under 7D diffeomorphisms, local 4D, $N=1$ superconformal transformations, and extensions thereof ({\it e.g.}\ extended supersymmetry transformations).
	\item The field strengths of the non-abelian tensor hierarchy (\ref{E:ATHXf}, \ref{E:KK}) are invariant under the linearized hierarchy transformations but not under the superconformal ones. So we make combinations that are invariant under one or two of these additional parameters using as few superspace derivatives as possible. 
	\item These partial invariants are used to covariantize (with respect to the superconformal parameters) the 3-form components we already had from the non-abelian tensor hierarchy.
This results in the eleven-dimensional 3-form superfield $\physical C_{\bm{abc}}$ \eqref{E:gauge3form} with leading Taylor component the physical 3-form in the on-shell spectrum.
	\item In the same way, we construct the eleven-dimensional frame superfield $\physical e_{\bm a}{}^{\bm b}$ \eqref{E:11Dframe} and gravitino superfield $\physical \psi_{\bm a}{}^{\bm \beta}$ \eqref{E:grino} from the partial invariants so that they transform as they should (\ref{E:translationParameter} and \ref{E:localSupersymmetry}). Taking curls defines the dimension-3/2 torsion $\FS{T}_{\bm {ab}}{}^{\bm \gamma}$ ({\it i.e.}\ the gravitino curvature)
and dimension-2 curvature $\FS{R}_{\bm {ab}}{}^{\bm {cd}}$.
	\item Finally, we use the partial invariants to covariantize the auxiliary fields in table \ref{T:Components}. The resulting invariants are listed in equation \ref{E:AuxiliarySuperfields}.
\end{enumerate}
These steps are summarized in table \ref{T:Summary}. 

\begin{table}[t]
{\footnotesize
\begin{align*}
{\renewcommand{\arraystretch}{1.5} 
\begin{array}{|ccccc|}
\hline
 ~~~\textrm{prepotential}  ~~~
 	& ~~~\textrm{partial invariant} ~~~ 
	& ~~~ \textrm{gauge field}  ~~~
	& ~~~ \textrm{auxiliary}  ~~~
	&  ~~~\textrm{field strength} ~~~\\
\hline
U^a 
	& \ast
	& \frameXX_a{}^b , \grinoXX_a{}^\beta
	& \dX^a 
	& \Torsion_{ab}{}^{\gamma}, \Curvature_{ab}{}^{cd} \\
\Psi_i^\alpha 
	& \X_i^a , \T_i 
	& \frameYX_i{}^a, \grinoXY_b{}^{\alpha i}
	&\l_i^\alpha , \y_i^a , \Ga_i^a,\t_{ab\,i},\r_i^\alpha  
	& \Torsion_{ab}{}^{\gamma k} , 
	\Curvature_{aj}{}^{cd} \\
\mathcal V^i 
	& \KK_\alpha^i
	& \frameXY_a{}^i, \grinoYX_i{}^{\beta}
	& \dY^i
	& \Curvature_{ab}{}^{ij} \\
X  	& \G , \S, \P
	& \CXXX_{abc}
	& \d, \R
	& \FSXXXX_{abcd}\\
\Sigma_i^\alpha 
	& \T_i 
	& \CXXY_{ab\, i}
	& \textrm{none}
	&\FSXYYY_{abc\, i}\\
V_{ij} 
	& \hatW^\alpha_{ij}
	& \CXYY_{a\, ij}, \spinorYY^\alpha_{ij} 
	& \dYY_{ij}
	&\FSXYYY_{ab\, ij} , \Curvature_{ab\, ij}\\
\Phi_{ijk}
	& \Z_{ijk} , \E_{ijkl}
	& \CYYY_{ijk}, \frameXX_i{}^j , \spinorYYY^\alpha_{ijk}
	& \m_{ijk} , \f_{ijk}
	& \FSXYYY_{a\, ijk}, \FSYYYY_{ijkl}, \Curvature_{aj\, kl}, \Curvature_{ij\, kl}\\
\hline
\end{array}
}
\end{align*}
\caption{\scriptsize Prepotentials and their derived quantities. All derived quantities contain admixtures of other prepotentials, so these identifications are approximate. 
(The assignment of curvature tensor components to any one field is similarly ambiguous.)
Explicit formul\ae{} for partial invariants in terms of the prepotentials are deferred to appendix \ref{S:Pregauge}.
(Hatted quantities are $\Psi$-corrected versions of the na\"ive hierarchy field strengths.)
The dimension-$1, \tfrac32, 2$ components of the field strength $\FS{G}$, torsion $\Torsion$, and curvature $\Curvature$ are the curls of the gauge fields as described in section \ref{S:Connections}.
}
\label{T:Summary}
} 
\end{table}

\subsection{Eleven-dimensional Superfields}
\label{S:11DSuperfields} 

In this section, we define superfields that have the property that their leading component corresponds to a field in eleven dimensions. 
These are either physical connections (and their curls) or auxiliary fields that do not propagate.

\subsubsection{Connection Superfields}
\label{S:Connections} 

We construct the superfields carrying the eleven-dimensional 3-form, frame, and gravitino as their leading component by combining the prepotentials \eqref{E:Prepotentials} and their derivatives into superfields that transform as connections. 
Much work goes into this construction in which successive combinations are made to have progressively simpler transformations under the gauge and local superconformal parameters. 
This is outlined in appendix \ref{S:Pregauge}, with the results presented here.

\paragraph{M-theory 3-form}
Under dimensional reduction, the eleven-dimensional 3-form splits up into a collection of $[p,q]$ forms embedded into superfields in \cite{Becker:2016xgv}. From this, we may extract the superfields for these components, and covariantize them with respect to the local superconformal transformations. 
This results in the following set of superfields:
\begin{subequations}
\label{E:gauge3form}
\begin{align}
\hyperlabel{E:CXXX}
\CXXX_{abc}&:= -\tfrac14\varepsilon_{abc\,{\delta \dt \delta} } \left(
	 [D^\delta, \bar D^{\dt \delta}] X 
	 	- (D^2 +\bar D^2) U^{\delta \dt \delta}
	\right)
\\
\hyperlabel{E:CXXY}
\CXXY_{\alpha \beta \, i}  &:= -\tfrac i2 D_{(\alpha} \left[ \,
	\Sigma_{\beta) i} 
	+\Psi_{\beta) i}\,
	\right]
\\
\hyperlabel{E:CXYY}
\CXYY_{\alpha \dt \alpha \, ij}&:= 
	\tfrac12 [D_\alpha, \bar D_{\dt \alpha}] V_{ij} 
	-\varphi_{ijk} \partial_{\alpha \dt \alpha} \mathcal V^k
	+ \varphi_{ijk} \X_{\alpha \dt \alpha}^k
\\
\hyperlabel{E:CYYY}
\CYYY_{ijk} &:= \tfrac12(\Phi_{ijk} + \bar \Phi_{ijk}) + \tfrac12 \psi_{ijkl} \T^l
~.
\end{align}
\end{subequations}
The combinations $\X_i^a$ and $\T_i$ may be thought of as partial covariantizations of the $\bar D\Psi$ and $D\Psi$ parts of the gravitino superfields (by the derivatives $\partial_i U^a$ and $\partial_i X$, respectively). 
They are defined explicitly in \eqref{E:X} and \eqref{E:T}, but we will not need that result.

Note, however that the corrections by $\{U^a, \Psi_i^\alpha, X\}$ to all the na\"ive components of the gauge 3-form are needed to cancel the local superconformal transformations of the compensating fields. 
With this, the corrected superfields transform as gauge $[p,q]$-forms, so we may immediately define the field strength superfield $\FS G_{\bm{abcd}} = 4 \partial_{[\bm a} \physical C_{\bm{bcd}]}$ corresponding to the M-theory 4-form invariant. 
(An explicit set of expressions for this in terms of tensor hierarchy invariants is given in \eqref{E:4formInvariants}.)

The construction given here is typical of everything that follows. We will now present the analogous results for the frame and gravitino.

\paragraph{Eleven-dimensional Frame}
The eleven-dimensional frame is determined by starting with the na\"ive term in the $\theta$-expansion \eqref{E:Prepotentials} of the prepotentials and adding terms so that the resulting superfields transform as $\delta \physical e_{\bm a}{}^{\bm b} = \partial_{\bm a} \xi^{\bm b} -\lambda_{\bm a}{}^{\bm b}$. 
Up to an $SO(3,1) \times SO(7)$ gauge choice and a normalization we fix by matching to reference \cite{Buchbinder:1998qv}, this fixes the components of the frame to be
\begin{subequations}
\label{E:11Dframe}
\begin{align}
\hyperlabel{E:frameXX}
\frameXX_{\un a}{}^{\un b} &:= 
	-\tfrac12 [D_\alpha, \bar D_{\dt \alpha} ] \, U^{\un b} 	
	-\tfrac23 \delta_\alpha^\beta \delta_{\dt \alpha}^{\dt \beta} \S
\\
\hyperlabel{E:frameXY}
\frameXY_{\un a}{}^{j} &:= \tfrac12 [D_\alpha, \bar D_{\dt \alpha} ] \, \mathcal V^j
\\
\hyperlabel{E:frameYX}
\frameYX_i{}^{\un b} &:= - \tfrac12 [
	\bar D^{\dt \beta} \Psi_i^\beta - D^\beta \bar \Psi_i^{\dt \beta} 
	]
\\
\hyperlabel{E:frameYY}
\frameYY_{i}{}^{j} &:= \tfrac14 \varphi^{j kl} \F_{ikl} 
	- \tfrac1{36} \delta_i^j \varphi^{klm} \F_{klm}
\end{align}
\end{subequations}
The translation and local Lorentz parameters are shared among these components, so they are also fixed. For the translations, this gives
\begin{align}
\label{E:translationParameter}
\xi^{\un a} :=-i ( \bar D^{\dt \alpha} L^\alpha + D^\alpha \bar L^{\dt \alpha} )
~~~\textrm{and}~~~
\xi^i &:= 
\tfrac1{2} (\tau^i +\bar \tau^i )
-\tfrac1{2i} (\Omega^i - \bar \Omega^i)
~.
\end{align}
We give the explicit expressions for the local Lorentz parameter \eqref{E:localLorentz} in appendix \ref{S:GrinoDescendants}.

We see explicitly from these expressions that the 4D component is as expected, and the Kaluza-Klein component is the na\"ive one. 
We also see that the real part of the complex vector in $\bar D \Psi$ is the ``lower-left block'' of the Kaluza-Klein decomposition. 
This block is usually taken to be zero, but in superspace that is only true to lowest order in the $\theta$-expansion. (The lowest component of $\frameYX_i{}^{b} $ vanishes in Wess-Zumino gauge. See \cite{Becker:2020hym} for a complete discussion of this component.)
We also see explicitly that the part $\frameYY_{i}{}^{j}$ of the linearized frame lying along the seven-dimensional space is a particular component of the partial field strength $\F_{ijk}$ for the ``scalars'' of the tensor hierarchy. (This is a partial covariantization of the imaginary part of the chiral field $\Phi_{ijk}$, see eq.\ \ref{E:F}.)

The symmetrization of this frame defines the linearized eleven-dimensional metric $\physical h_{\bm {ab}} : = 2\physical e_{(\bm {a}}{}^{\bm c}\delta_{\bm {b})\bm {c}}$.
The Riemann tensor $\FS R_{\bm {ab}}{}^{\bm {cd}} :=  \partial_{[\bm {a}} \partial^{[\bm {c}} h_{\bm {b}]}^{\bm {d}]}$ is the double curl of this.
The linearized curvature 2-form $\FS R_{\bm {ab}}{}^{\bm {cd}} := 2 \partial_{[\bm {a}} \physical \omega_{\bm {b}]}{}^{\bm {cd}}$ is the curl of the spin connection. 
These curvatures agree component-wise when the spin connection is taken to be torsion-free, since $\FS T_{\bm {ab}}{}^{\bm c} = 0$ implies $\omega_{\bm {c}}{}^{\bm {ab}} := \tfrac12 \partial^{[\bm {a}} \physical h_{\bm {c}}^{\bm {b}]}$.

\paragraph{Eleven-dimensional Gravitino}
The definition of the eleven-dimensional gravitino suffers from various ambiguities. 
Below, we present what we consider the simplest one. 
To do so, it is helpful to first define the components (which appear also in table \ref{T:Components})
\begin{subequations}
\label{E:spinor}
\begin{align}
\hyperlabel{E:spinorYY}
\spinorYY_{ij}^\alpha &:=
	\W_{ij}^\alpha +2\partial_{[i} \Psi_{j]}^\alpha
	+i\varphi_{ijk} (\KK^{\alpha k} -\tfrac14 \bar D^2 \Psi^{\alpha k})
\\
\hyperlabel{E:spinorYYY}
\spinorYYY_{ijk}^\alpha
	&:= D^\alpha \left( i\F_{ijk} 
		+\tfrac 12 \psi_{ijkl}  \T^l 
		+3 \varphi_{l[ij} \partial_{k]} \mathcal V^l\right)
~.
\end{align}
\end{subequations}
The first of these may be understood as the covariantization (canceling the local superconformal shifts) of the 21 gaugini in the abelian vector multiplet prepotential $V_{ij}$ by the Kaluza-Klein gauge field $\mathcal V^i$ and the gravitino superfield $\Psi_i^\alpha$.
The second is the analogous thing for the 35 spin-1/2 fields in the chiral multiplet $\Phi_{ijk}$. (The part of $\Psi_i^\alpha$ that enters 
requires the further covariantization by $\Sigma_i^\alpha$ into $\T$.)
Together, they comprise 56 spin-1/2 fields from the 4D perspective \cite{Cremmer:1978ds}.

Using various $G_2$ projectors given in appendix \ref{S:G2}, these components can be reorganized into the $56 = 7+ 49$ components $ \physical \psi_i{}^\beta, \physical \psi_i{}^{\beta j}$ of the linearized eleven-dimensional gravitino. Using this, we explicitly define the eleven-dimensional gravitino components as
\begin{subequations}
\label{E:grino}
\begin{align}
\hyperlabel{E:grinoXX}
\grinoXX_{\un a}{}^\beta &:=-\tfrac{i}8 \bar D^2 D^{\beta} U_{\un a}
	+\tfrac i{3} \delta_\alpha^\beta
		\bar D_{\dt \alpha} \S
\\
\hyperlabel{E:grinoYX}
\grinoYX_i{}^\beta 
	&:= \tfrac1{12}\l_i^\beta-\tfrac1{12}\left[
	\varphi_i{}^{jk} \spinorYY^\beta_{jk}
	-\tfrac 1{6} \psi_i{}^{jkl} \spinorYYY^\beta_{jkl}
	\right]
\\
\hyperlabel{E:grinoXY}
\grinoXY_{\un a}{}^{\beta j} 
&:= 2 D^\beta \X_{\un a}^j
	-\tfrac23 \delta_\alpha^\beta \left[
		2D^\gamma \X^j_{\gamma \dt \alpha} 
		+ \bar D_{\dt \alpha} \T^j 
		+2i \bar \KK^j_{\dt \alpha}
	\right]
\\
\hyperlabel{E:grinoYY}
\grinoYY_i{}^{\beta j} 
	&:=
	\hmap_i{}^j(\spinorYYY^\beta_{klm})
	-\tfrac 1{36} \varphi_i{}^{jk} \psi_k{}^{	lmn} \spinorYYY_{lmn}^\beta
	-\pi_{\bm{14}} ( \spinorYY^\beta_i{}^j )
\end{align}
\end{subequations}
(In the last component, we see explicitly the $49 = 28  + 7 + 14$ structure (recall eq.\ \ref{E:hmap}) and how the original $\bm {21}$ and $\bm {35}$ are mixed together through $G_2$ representations to form the $\bm 7$ and $\bm {49}$.)

These components are defined by their gauge transformations 
\begin{align}
\label{E:localSupersymmetry}
\delta \physical \psi_{\bm a}{}^{\bm \beta} = \partial_{\bm a} \epsilon^{\bm \beta}
~~~\textrm{with}~~~
\epsilon^\beta := -\tfrac14 \bar D^2 L^\beta
~~~,~~~
\epsilon^{\beta j} := - 2 D^\beta \Omega^j 
~,
\end{align}
so that their curls $\FS{T}_{\bm {ab}}{}^{\bm \gamma}$ are invariant; they are the dimension-3/2 torsion components.
But this condition is ambiguous: As we will explain presently (cf.\ eq.\ \ref{E:grinEoM}), there is an invariant spinor superfield $\l^i_\alpha$ of dimension 1/2 that can be added to three of the four components defined above. 
The ones we have defined here are as we originally derived them. Curiously, they also turn out to be the simplest ones in the sense that they imply that all the dimension-1/2 torsions vanish, as we will see in section \ref{S:Supergeometry}.

\subsubsection{Auxiliary Superfields}
\label{S:Auxiliary}

We will now define superfields containing the auxiliary components as the leading term of their $\theta$-expansion. 
These ``auxiliary superfields'' will be given the same name as their leading component to avoid a further proliferation of new notation. 
This should cause no confusion since, throughout this paper, we will always mean the superfield unless we explicitly state otherwise. 

The definition of the dimension-1/2 gravitini was ambiguous because there is a dimension-1/2 invariant
\begin{subequations}
\label{E:AuxiliarySuperfields}
\begin{align}
\hyperlabel{E:grinEoM}
\l_i^\alpha
	&:= 2\bar D_{\dt \alpha} \X_i^{ \un a} 
	-D^\alpha \T_i 
	+2i \J_i^\alpha 
~.	
\end{align}
This field is present in any Lagrangian theory of the gravitino multiplet $\Psi_i^\alpha$ where, as a component field, it forms a Lagrange multiplier pair with a dimension-3/2 spinor $\r_i^\alpha$ to be defined below. 
Because of this, and as we will see explicitly in section \ref{S:Dynamics}, this superfield is proportional to the $\Psi_i^\alpha$ equation of motion.

There are many more dimension-1 auxiliary fields since most 4D, $N=1$ superfields contain one. 
The vector multiplets all contain a real auxiliary field we denote by $\auxiliary d$. There are 4 such real prepotentials: the conformal graviton $U^a$, the Kaluza-Klein field $\mathcal V^i$, the 3-form $X$, and the abelian vectors $V_{ij}$. Their $\auxiliary d$-components are covariantized as 
\begin{align}
\hyperlabel{E:dX}
\dX^{\un a} &:=\tfrac18 D^\beta \bar D^2 D_\beta U^{\un a} 
	+ \tfrac16 [ D^\alpha, \bar D^{\dt \alpha}] \S
	+\partial^{\un a} \P
\\
\hyperlabel{E:d}
\d
&:=
-\tfrac 14 (D^2 + \bar D^2) \S
\\
\hyperlabel{E:dY}
\dY^i
	&:=\tfrac13 D^\alpha \KK_\alpha^i 
		-\tfrac23 \partial^{a} \X_{a}^i
		+\tfrac i{12} (D^2 - \bar D^2) \T^i
\\
\hyperlabel{E:dYY}
\dYY_{ij}&:= -\tfrac1{32} \psi_{ij}{}^{kl} \big[
		D^\alpha \spinorYY_{\alpha \, kl}
		+\bar D_{\dt \alpha} \bar \spinorYY_{kl}^{\dt \alpha }
	\big]
	-\tfrac16 \varphi_{ijk} \partial^k \S
	+\tfrac14 \hitchin_{ijk\, lmn}\partial^k  \F^{lmn}
~.
\end{align}

Another familiar 4D, $N=1$ auxiliary is that of the chiral scalar field, of which we have only one such representation $\Phi_{ijk}$.
Covariantizing its auxiliary $\auxiliary f$-component, we find
\begin{align}
\hyperlabel{E:f}
\f_{ijk}	
	&:= D^2 \Z_{ijk} 
		+\tfrac i3 (\hitchin^{-1})_{ijk\, lmn}\varepsilon^{lmnpqrs}\partial_p \Z_{qrs}
~.
\end{align}
(Here $\Z_{ijk}$ is the partial covariantization of the scalars $\F_{ijk}$ defined in \eqref{E:Z}.)

Finally, there are the additional auxiliary fields of the gravitino multiplet. 
This multiplet is reviewed in detail in appendix C of \cite{Becker:2017zwe}. 
In addition to the dimension-1/2 auxiliary field \eqref{E:grinEoM}, there are two dimension-1 components 
\begin{align}
\hyperlabel{E:y}
	\y^i_{\alpha \dt \alpha}&:= - \bar D_{\dt \alpha} \l_\alpha^i
\\
\hyperlabel{E:t} 
\t_{\alpha \beta i} &:=\tfrac12 D_{(\alpha} \KK_{\beta) i}
	-\tfrac i4 \big[  D_{(\beta} \bar D^{\dt \alpha}
	+3 \bar D^{\dt \alpha} D_{(\beta}   
	\big] \X_{\alpha) \dt \alpha i}
\cr&=
	-\tfrac14  \bar D^2 D_{(\alpha} \Psi_{\beta) i}
	+\tfrac i4 \partial_i \big[  D_{(\beta} \bar D^{\dt \alpha}
	+3 \bar D^{\dt \alpha} D_{(\beta}    \big] U_{\alpha)\dt \alpha} 
	+ \tfrac i2 \partial_{(\beta}{}^{\dt \alpha} \gravitonXY_{\alpha)\dt \alpha 
i}	
~,
\end{align}
comprising a complex vector and a 2-form, 
and the top component of $\Psi$ 
\begin{align}
\hyperlabel{E:r}
\r^i_\alpha&:=-\tfrac14 D^\beta \bar D^2 D_{(\alpha} \Psi_{\beta)}^i 
	- i \partial^i \big[
	D_\alpha \G +\tfrac12 \bar D^{\dt \alpha} D^2 U_{\un a}
	\big]
~,
\end{align}
\end{subequations}
completing the set of auxiliary fields with a dimension-3/2 spinor.

The auxiliary fields defined in this section are important because they are part of a linear representation of the subalgebra generated by 4 of the 32 supersymmetries. 
Many of them appear explicitly in the supersymmetry transformation \eqref{E:SupersymmetryTransformationsGravitino} of the eleven-dimensional gravitino. 
To compare this rule to the known on-shell transformation \eqref{E:LinSUSY} and to prove that they do not contribute propagating degrees of freedom, we examine in section \ref{S:Dynamics} how they appear in the off-shell dynamics. 

\paragraph{Useful Alternative Composite Fields}
The supersymmetry transformations of the gravitino are mixtures of the auxiliary fields and 4-form flux. 
The gauge-invariant combinations that appear naturally in the differential algebra are
\begin{subequations}
\begin{align}
\hyperlabel{E:R}
\R &:= -\tfrac 16 \bar D^2 \S = \tfrac i6 \bar D^2 \P 
	= -\tfrac 1{12} \bar D^2 [ \bar \G + i \partial_{\un a} U^{\un a} ]
	= \tfrac13 \d + \tfrac i{6\cdot 4!} \varepsilon^{abcd} \FSXXXX_{abcd}
\\
\hyperlabel{E:Ga}
\Ga^{a}_i &:= -i\bar D^2 \X^{a}_i
	+\tfrac i2 \bar D \bar \sigma^a D \T_i 
\cr
	&=  \tfrac 1{2i} \big[ \y_i^{a} - \bar \y_i^{a} \big]
	+\tfrac16 \psi_{ijkl} \FSXYYY^{a jkl} 
	+i \tilde \FSXXXY_i^{a}
\\
\hyperlabel{E:m}
\m^{ijk}
	&:= \hitchin^{ijk\, lmn} D^2  \Z_{lmn} 
	- \tfrac 13 \epsilon^{ijk lmnp}\partial_l \bar \Phi_{mnp}
	= (\hitchin)^{ijk\, lmn}  \f_{lmn}-\tfrac 1{12}  \varepsilon^{ijklmnp} \FSYYYY_{lmnp}
\\
\hyperlabel{E:Ztorsion}
\Ztorsion_i{}^j 
	&:= 
	- i \hmap_i{}^j ( \f_{lmn})
	+\tfrac i{36}\varphi_i{}^{jk}\psi_k{}^{lmn} \f_{lmn}
	-8  \pi_{\bm{14}} \dYY_i{}^j	
\end{align}
\end{subequations}
The first of these is the chiral scalar field strength of old-minimal supergravity containing the complex auxiliary field, the spin-1/2 part of the gravitino, and the curvature scalar \cite{Gates:1983nr,Wess:1992cp,Buchbinder:1998qv}.
In the 3-form formulation of the theory employed here, the imaginary part of the auxiliary field is dualized to the 4-form component $\FSXXXX_{abcd}$ \cite{Gates:1980ay, Ovrut:1997ur}.

The second is a complex linear field $\bar D^{(\dt \beta} \Ga^{\dt \alpha)\alpha}_i = 0$. In fact, $\Ga_i^{\dt \alpha \alpha} = {12}i \bar D^{\dt \alpha} \grinoYX_i^\alpha$, which explains why it is this combination that features prominently in the gravitino supersymmetry transformations.
Expanding it out explicitly, we see that it is related to the imaginary part of the gravitino auxiliary field $\y_i^a$. (Note, however, that the real part of $\y$ does appear in the transformation of $\grinoXY_a{}^{\beta j}$ \eqref{E:SupersymmetryTransformationsGravitino} cf.\ also eq.\  \ref{E:DgrinoXY}.)

The third is a modification of the auxiliary field of the scalars \eqref{E:f} by a real field strength. 
It is an interesting alternative that is chiral.\footnote{
Contraction with the constant $\varepsilon_{i_1\dots i_7}$ symbol defines a chiral 4-form 
\begin{align*}
\bm E_{ijkl} 
&:= -\tfrac 1{12} \varepsilon_{ijklmnp} \bar \m^{mnp}
	= 4 \partial_{[i} \Phi_{jkl]} 
	- \tfrac1{12} \varepsilon_{ijklmnp} \bar D^2 \left[ \hitchin^{mnp\,qrs} \bar \Z_{qrs}\right]
~.
\end{align*}
} 

The last combination $\Ztorsion_i{}^j$ of auxiliary fields appears in the supersymmetry transformation of $\grinoYY_i{}^{\beta j}$ (cf.\ eq.\ \ref{E:SupersymmetryTransformationsGravitino}).
Equivalently, this final invariant is the component of the dimension-1 torsion $\Torsion_{\alpha j}{}^{\gamma k} =4 \delta_\alpha^\gamma \Ztorsion_j{}^k +\cdots $ that is a singlet under the 4D part of the Lorentz group (see eq.\ \ref{E:TorsionY}).
In the form presented, the decomposition $\bm {49} = \bm {28} + \bm {7} + \bm {14}$ is evident. 
Of these the $\bm {28} = \bm {27} +\bm {1}$ and $\bm 7$ are complex, whereas the $\bm {14}$ is real.
This agrees with the counting of the first-order invariants of the $G_2$ structure, with the real $\bm {49} = \bm {27} + \bm {1} + \bm {7}+\bm {14}$ corresponding to the four torsion forms \eqref{E:TorsionForms} and the imaginary $\bm {35}= \bm {27} + \bm {1}+\bm {7}$ to 
the analogous derivatives of the gauge 3-form (the derivative in the $\bm{14}$  is not gauge invariant).

Finally, we mention for completeness that additional invariants are possible that look new, but are actually linear combinations of the invariants defined above.
For example, it is easy to check that the dimension-3/2 spinor $\bar D^2 D^\alpha \T_i$ is invariant. It is somewhat less trivial to show that it satisfies
\begin{align}
\tfrac {3i}{4} \bar D^2 D_\alpha \T_i
	&= \r_{\alpha i} 
	+ 2 (\sigma^{ab})_\alpha{}^\beta \partial_{[a}\grinoXX_{b] \beta i}
	-12 (\sigma^a)_{\alpha \dt \alpha} 
		[
		\partial_a \bar \grinoYX_i{}^{\dt \alpha}
		-\partial_i \bar \grinoXX_a{}^{\dt \alpha}
		]
~.		
\end{align}
We conclude that this invariant is not new, since it is a combination of 
the previously-defined auxiliary field $\r_i^\alpha$ \eqref{E:r} and curls of the gravitini \eqref{E:grinoXX} and \eqref{E:grinoYX}.
(The choice of the dimension-3/2 spinor depends on how we order the derivatives $D$ and $\bar D$ in the definition of the auxiliary component.)
This conclusion was guaranteed by the component results of \cite{Becker:2017zwe} where a Wess-Zumino gauge analysis was carried out to identify the spectrum: Any new invariant would have to correspond to a physical component field, but this would be in contradiction with the Wess-Zumino analysis since we already accounted for all of those fields.

\section{Dynamics}
\label{S:Dynamics}
The prepotential superfields of tabel \ref{T:Components} are not constrained except for reality and chirality conditions. 
Therefore, we can write an action in terms of them to determine the dynamical equations. 
This was done at the linearized level in \cite{Becker:2017zwe} (and to low order in a gravitino superfield expansion in \cite{Becker:2018phr}).\footnote{The gravitino and Kaluza-Klein field were treated non-linearly in reference \cite{Becker:2020hym} by including them into the superspace geometry. In this approach $\Psi_i^\alpha$ and $\mathcal V^i$ all appear in the connections alongside $U^a$.} 
Defining the variation ${\color{Red}\bm E_\phi} := \delta S/\delta \phi$ of that action with respect to the prepotential $\phi$, we obtain the supercurrents
\begin{subequations} 
\label{E:SuperEoM}
\begin{align}
\hyperlabel{E:EoMX}
\EoMX{}_a 	
&= - \dX_a +\tfrac1{18} \varphi^{ijk} \FSXYYY_{a\, ijk}
\\
\hyperlabel{E:EoMgrino}
\EoMgrino{}^i_\alpha &= \tfrac i4 \l_\alpha^i
\\
\hyperlabel{E:EoMY}
\EoMY{}_i 
	&= - 3\dY_i
	 +\tfrac16\varphi^{jkl} \FSYYYY_{ijkl}
\\
\hyperlabel{E:EoM}
\EoM 
	&= -\tfrac23 \d
	+\tfrac 1{72}\varphi_{ijk}\hitchin^{ijk\, lmn}  (\f_{lmn} +\bar \f_{lmn})
	+\tfrac 1{48}\psi^{ijkl}\FSYYYY_{ijkl}
\cr
	&= -\tfrac23 \d
	-\tfrac 1{96}\varphi_{ijk}(\m^{ijk} +\bar \m^{ijk})
	+\tfrac 1{96}\psi^{ijkl}\FSYYYY_{ijkl}
\\
\label{E:EoMSigma}
\vanishing{E_{\Sigma} }{}_\alpha^i &= 
	\tfrac i{16}\bar D^2 \l_\alpha^i
\\
\hyperlabel{E:EoMYY}
\EoMYY{}^{ij}
	&= 2\dYY^{ij} 
\\
\hyperlabel{E:EoMYYY}
\EoMYYY^{ijk}	
	&= 
	\tfrac i{48} \hitchin^{ijk\, lmn}  \bar \f_{lmn}
	-\tfrac i{576} \varepsilon^{ijklmnp} \FSYYYY_{lmnp}
	+\tfrac i6 \varphi^{ijk} \R
\cr
	&= 
	\tfrac i{48} \bar \m^{ijk} 
	+\tfrac i{12} \varphi^{ijk} \R
~.
\end{align}
\end{subequations}
These are in bold font because they are gauge-invariant, and we have colored them red to indicate that setting them to zero defines the superspace mass-shell condition.\footnote{This overloading of notion should not be confused with the compensating component fields, which are unrelated and have played no role since the earliest part of section \ref{S:Prepotentials}.}
This is a useful distinction because, although {\em we will continue to work off-shell}, these equations are needed to prove the claim that auxiliary fields do not contribute any new degrees of freedom on-shell: The auxiliary fields differ from components of the physical 4-form field strength only by the superfields $\color{Red}\bm E_\ast$ that vanish on-shell.

Explicitly, we can solve these equations \eqref{E:SuperEoM} for the auxiliary fields to find
\begin{subequations}
\begin{align}
\dX_a &=\tfrac1{18} \varphi^{ijk} \FSXYYY_{a\, ijk} 
	- \EoMX{}_a
\\
\l_\alpha^i &= -4i \EoMgrino{}_\alpha^i
\\
3\dY_i&= \tfrac16\varphi^{jkl} \FSYYYY_{ijkl}
	 - \EoMY{}_i
\\
\tfrac12 \d
&= 
	-\tfrac 1{96}\psi^{ijkl}\FSYYYY_{ijkl}
	+\EoM
	-\tfrac i2 \varphi_{ijk} ( \EoMYYY - \overline {\EoMYYY} )^{ijk}
\\
\dYY_{ij}  &= \tfrac12 \EoMYY{}_{kl} 
\\
\hitchin^{ijk\, lmn} \f_{lmn}  &= 
\tfrac 1{36}\left[
	\varphi^{ijk}(i  \varepsilon^{abcd} \FSXXXX_{abcd} 
	+  \psi^{lmnp} \FSYYYY_{lmnp} )
	+3 \varepsilon^{ijklmnp} \FSYYYY_{lmnp} 
	\right]
\cr
	&\hspace{20mm}
	+48i\overline\EoMYYY^{ijk}
	+\tfrac{4i}3 \varphi^{ijk} \varphi_{lmn} ( \EoMYYY - \overline {\EoMYYY})^{lmn}
	-\tfrac 83\varphi^{ijk} \EoM 
\end{align}
\end{subequations}
The auxiliary fields not appearing here $\{ \y_i^a, \t_i^{\alpha \beta}, \r_i^\alpha \}$ are those of the gravitino superfield. 
However, it is not difficult to show that
\begin{align}
\y_{i}^a&= -2i \bar D \bar \sigma^a \EoMgrino{}_i
\cr
\t_{\alpha \beta i} &=
		-\tfrac12 \varphi_i{}^{jk} \FSXXYY_{\alpha \beta jk} 
		- D_{(\alpha}\EoMgrino{}_{\beta) i}
\cr
\r_i^\alpha 
	&=
	-\tfrac 1{12} (\bar D\bar \sigma^a)^{\alpha} 
		\left[ 
			\psi_i{}^{jkl} \FSXYYY_{a\, jkl} 
			-i \varepsilon_a{}^{bcd}  \FSXXXY_{bcd\, i}
		\right]
	+2i \psi_i{}^{jkl} \partial_j \spinorYY_{kl}^\alpha 
	- i \bar D_{\dt \alpha} D^\alpha \overline \EoMgrino^{\dt \alpha}_{ i} 
~.	
\end{align}
It follows that $\y$ and $\t$ contribute nothing new on-shell. 
Similarly, up to equations of motion, $\r$ is a sum of three curls, each of which is separately an invariant of dimension 3/2---that is---they are components of the gravitino curl. 

Finally, we note that the action of the superspace derivatives on the field strength $\FS{G}=d\physical{C}$ ({\it i.e.}\ its supersymmetry transformation) generate gravitino curls by \eqref{E:SupersymmetryTransformationsThreeform}, up to supersymmetry spacetime derivatives of $\sim D\l_i $. 
The latter can be rewritten in terms auxiliary fields and field strengths by \eqref{E:y} and 
\begin{subequations}
\begin{align}
D_{(\alpha} \l_{\beta)}^i & = 4i \t_{\alpha \beta}{}^i + 2i \varphi^{ijk} \FSXXYY_{\alpha \beta jk} 
\\
D^\alpha \l_{\alpha}^i &= 
	6i \dY^i 
	+  4\varphi^{ijk} \dYY_{jk} 
	-\tfrac i6 \psi^{ijkl} \f_{jkl}
~,	
\end{align}
\end{subequations}
implying that supersymmetric derivatives of the dimension-1 field strength generate no new invariants. 
This concludes the demonstration that the $N=1/8$ off-shell component spectrum reduces correctly to the physical spectrum if the component equations of motion are imposed by taking $\color{Red}\bm E_\ast| \to 0$.

\section{Toward Eleven-dimensional Off-shell Supergeometry}
\label{S:Supergeometry}

The analysis that leads to the results of the previous sections also implies the constraints on the supergeometry needed to solve the superspace Bianchi identities. 
Such a top-down approach defines the superspace torsion and curvature through the commutator 
\begin{align}
\label{E:TorsionCurvature11D}
[ \nabla_{\bm A}, \nabla_{\bm B} ] = 
	T_{\bm A \bm B}{}^{\bm C} \nabla_{\bm C} 
	+\tfrac12 R_{\bm A \bm B}{}^{\bm c\bm d} M_{\bm c \bm d} 
~,	
\end{align}
where $\nabla_{\bm A} = \nabla_{\bm \alpha} , \nabla_{\bm a}$ is the covariant superspace derivative of eleven-dimensional supergravity (${}_{\bm \alpha} = {}_1,\dots,{}_{32}$ and ${}_{\bm a} = {}_0,\dots , {}_{10}$). 
These are subject to the Bianchi identities that follow from the nested commutator identity 
\begin{align}
\label{E:Jacobi11D}
[[ \nabla_{\bm A}, \nabla_{\bm B} ] , \nabla_{\bm C} ] ] + \textrm{permutations}= 0
\end{align}
The non-trivial part says that 
\begin{align}
\label{E:BI11D}
\nabla_{[\bm C} T_{\bm A \bm B]}{}^{\bm D} 
+T_{[\bm A \bm B]}{}^{\bm E} T_{|\bm E| \bm C] }{}^{\bm D} 
	+\tfrac12 R_{\bm A \bm B}{}^{\bm c\bm d} (\Gamma_{\bm c \bm d})_{\bm C}{}^{\bm D} = 0
\end{align}

These identities become non-trivial when covariant conditions are imposed on various components of the torsion and curvature. Some of these conditions are a matter of convention that follows from the possibility of shifting the definitions of the various connections (most importantly, the spin connection).\footnote{
This can be formalized in terms of first-order-flat $G$ structures \cite{Lott:1990zh, Lott:2001st}. 
The obstruction theory for this is known as Spencer cohomology (cf.\ {\it e.g.}\ Schwachh\"ofer's article ``Holonomy Groups and Algebras'' in \cite{book:976236}). The Spencer cohomology for 11D supergravity was considered in \cite{Figueroa-OFarrill:2015rfh}.} 
For eleven-dimensional supergravity in superspace, these are
\begin{align}
\label{E:TorsionConstraints11D}
T_{\bm \alpha \bm \beta}{}^{\bm \gamma} = 0
~~~,~~~
T_{\bm \alpha \bm \beta}{}^{\bm c} = -2i (\Gamma^{\bm c} )_{\bm \alpha \bm \beta}
~~~,~~~
T_{\bm a\bm b}{}^{\bm c} = 0
~.
\end{align}
Others follow from the deformation of constraints defining representations of the flat space supersymmetry algebra (most notably, chiral).
There is no known analog of this type of constraint in eleven-dimensional supergravity, presumably because there are no matter multiplets with spins $ < 3/2$ in this case. 

Historically, appropriate torsion constraints were often found by making an {\it Ansatz} and solving the Bianchi identities. 
Poor {\it Ans\"atze} over-constrain the torsions leading to trivial solutions or on-shell geometries. We can circumvent this problem by reducing the structure group so that there are fewer torsion constraints and more possibilities for deformed representation constraints. 
Of course, guessing the torsion constraints is complicated by the lack of symmetry.

But now we notice that we already know the solution to the linearized constraints, even though we do not know the constraints!
This is because in the case of off-shell supergeometry, the solution to the superspace Bianchi identities culminates in explicit expressions for the field strengths, torsions, and curvatures in terms of unconstrained prepotentials. We already know from the derivation of the correct spectrum \cite{Becker:2017zwe}, that these must be equivalent to the prepotentials \eqref{E:Prepotentials}.
Starting with these, then, it must be possible to derive the constraints from this solution directly. 
In principle, this can be done by acting repeatedly on the prepotentials with superspace derivatives in all possible ways to generate a graded space of superfields. 
In this large space, we will find (among many other things) invariants and relations between them. 
The invariants are the aforementioned field strengths, torsions, curvatures, and their derivatives, since these are the only invariants of a superspace by definition. 
The relations between them will then be the (reduced) Bianchi identities.
In this paper, we have taken a shortcut by building the connections and covariantizing the auxiliary fields. The remaining invariants are the curls of the connections. The constraints can then be gleaned from the spectrum of invariants and the supersymmetry transformation of the connections.

\paragraph{Off-shell 11D, $N=1/8$ Supergeometry }
Let us now specialize to the supergeometry appropriate to the eleven-dimensional superspace considered in this paper.
Since only $N=1/8$ is manifest, we retain from the superspace covariant derivatives above only $\{\nabla_\alpha, \bar \nabla_{\dt \alpha}, \nabla_a, \nabla_i\}$. Similarly, we keep only those parts of the Bianchi identity \eqref{E:BI11D} with the implied free indices.
However, the on-shell eleven-dimensional torsion \eqref{E:TorsionConstraints11D} is a tensor, so any non-vanishing component of it must be retained including $T_{\alpha\, \beta j}{}^k = -2i \varepsilon_{\alpha \beta} \delta_j^k $.
The implication of this is that torsion components of the form $T_{\alpha \beta j}{}^{\ast}$ can appear through the quadratic terms in \eqref{E:BI11D}.
The remaining dimension-0 constraints of relevance from \eqref{E:TorsionConstraints11D}  are
\begin{align}
T_{\alpha \beta}{}^c = 0
~,~~
T_{\alpha \dt \beta}{}^c = -2i (\sigma^c)_{\alpha \dt \beta} 
~,~~~
T_{\alpha \dt \beta}{}^k = 0
~
\end{align}
together with their conjugates. 
This is consistent with our finding that there are no invariants with dimension $<1/2$. 

Less obvious---but not difficult to verify---is that there are also no dimension-1/2 torsions with our choice of gravitino components: 
In this form, the gravitino auxiliary field appears only in the dimension-1/2 component 
\begin{align}
\FS G_{\alpha b jk} = -\tfrac16 (\sigma_b \bar \l^i)_\alpha \varphi_{ijk} 
\end{align}
of the super-4-form field strength.

This leaves the dimension-1 conventional constraints. 
Since there is no bosonic eleven-dimensional torsion, we have chosen the constraint accordingly: $T_{\bm{ab}}{}^{\bm c} =0$ defines the 11D spin connection in terms of the frame as we have chosen it in section \ref{S:11DSuperfields}.

We could now begin to solve the superspace Bianchi identities subject to these conditions. 
This would be useful as a direct method of deriving the field strengths, perhaps even beyond their linearized approximation.
Instead, we observe that the dimension-1 torsion components can be deduced directly from the linearized gravitino transformations 
\begin{gather}
\delta_\epsilon \psi_{\bm a}{}^{\bm \beta} 
	\sim \delta_\epsilon E_{\bm a}{}^{\bm \beta}
	\sim -\epsilon D E_{\bm a}{}^{\bm \beta} 
	\sim -\epsilon^{\gamma} \Torsion_{\gamma \bm a}{}^{\bm \beta} 
		-\epsilon^{\dt \gamma} \Torsion_{\dt \gamma \bm a}{}^{\bm \beta} 
~.
\end{gather}
Collecting the relevant parts from \eqref{E:SupersymmetryTransformationsGravitino} (or app. \ref{S:GrinoDescendants}), we find for the $\Torsion^{\gamma}$ part
\begin{gather}
\label{E:TorsionX}
\Torsion_{\alpha b}{}^{\gamma} = 
	 \tfrac i2  [\delta^\gamma_\alpha \eta_{bc} + (\sigma_{bc})_\alpha {}^{\gamma}] \dX^c
~,~~
\Torsion_{\alpha b}{}^{\dt \gamma} = 
	\tfrac 12 (\sigma_b)_{\alpha}{}^{\dt \gamma} \R
~,~~
	\cr	
\Torsion_{\alpha j}{}^{\gamma} = 
	\tfrac i2 \delta_\alpha^\gamma \dY_j 
	+\tfrac i{12} (\sigma^{ab})_\alpha {}^\gamma \t_{abj}
~,~~
\Torsion_{\alpha j}{}^{\dt \gamma} = \tfrac i{12} (\sigma_a)_\alpha{}^{\dt \gamma} \bar \Ga_j^a
~.~~
\end{gather}
The first line is the same as those of textbook treatments of 4D, $N=1$ superspace \cite{Gates:1983nr,Wess:1992cp,Buchbinder:1998qv}.
They are essentially the definitions of the $\dX^a$ and $\R$ auxiliary fields that contain the 4D part of the Ricci tensor and scalar curvature in their $\theta$-expansion.
We see that for D $>4$, these are extended by $\dY_i$, $\t_{ab \,k}$, and $\bar \Ga_i^a$.

The analogous dimension-1 torsion $\Torsion^{\gamma k}$ with $\bm 7$-valued spinor index decomposes as
\begin{align}
\label{E:TorsionY}
\Torsion_{\alpha b}{}^{\gamma k} &=
	\tfrac i{12} [5 \delta_\alpha^\gamma \eta_{bc} + 8 (\sigma_{bc})_\alpha {}^\gamma ] \bar \Ga^{ck} 
	+ \tfrac i8 \delta_\alpha^\gamma \Ga_b^k 
	-\tfrac 18 \delta_\alpha^\gamma 
		[ \bar D \bar \sigma_b \l^k + D\sigma_b \bar \l{}^k]
~,~~
\cr
\Torsion_{\alpha b}{}^{\dt \gamma k} &=
	i(\sigma_b)_\alpha{}^{\dt \gamma} \dY^k 
	+\tfrac i6 (4 \eta_{bc} \sigma_d - i \epsilon_{bcde} \sigma^e)_\alpha {}^{\dt \gamma} \t^{kcd} ,
~,~~
	\cr	
\Torsion_{\alpha j}{}^{\gamma k} &=
	\tfrac12 \delta_\alpha^\gamma \Ztorsion_j{}^k
	+ i (\sigma^{ab})_\alpha{}^\gamma \pi_{\bm {14}}\FSXXYY_{ab\, j}{}^k 
~,~~
\cr
\Torsion_{\alpha j}{}^{\dt \gamma k} &= 
	 i (\sigma^a)_\alpha{}^{\dt \gamma} [
		\hmap_j{}^k(\FSXYYY_{a\,lmn}) 
	-\tfrac 1{6} \varphi_j{}^{kl} \bar \Ga_{a l}
	]
	- \tfrac 16 \varphi_j{}^{kl} D_\alpha \bar \l^{\dt \gamma}_l
~.~~	
\end{align}
The new structures include the real and imaginary parts of the descendant $\y \sim \bar D\l$ of the gravitino auxiliary (which vanish on-shell), various projections of the 4-form field strength, and the superspace covariantization $\Ztorsion_i{}^j$ of the $G_2$ torsion from \eqref{E:Ztorsion}.

The remaining torsion component was defined already at the end of section \ref{S:Connections} as the spacetime curl of the eleven-dimensional gravitino.
The dimension-2 invariants (curvature) were also defined in that section.
In summary, we have demonstrated that the only field strengths of the eleven-dimensional supergeometry with 4 off-shell supersymmetries are the dimension 1 4-form $\FS G_{\bm{abcd}}$, the dimension-3/2 torsion $\Torsion_{\bm {ab}}{}^{\bm \gamma}$, and the dimension-2 curvature $\Curvature_{\bm{ab}}{}^{\bm{cd}}$. These are all superfields with the implied component field strengths as their leading Taylor coefficient. 
In addition, we find the auxiliary superfields \eqref{E:AuxiliarySuperfields} with $1/2\leq$ dimension $\leq 3/2$, which are reduced dynamically to the aforementioned physical spectrum by setting the supercurrents \eqref{E:SuperEoM} to zero.

\section{Epilogue}
\label{S:Epilogue}

We have built up the linearized tensors of eleven-dimensional supergravity directly from the unconstrained prepotentials of this theory with the structure group reduced to $SO(3,1) \times G_2$.
This gives a superfield for each physical component field of eleven-dimensional supergravity. 
In addition, the construction provides superfields embedding the auxiliary fields needed for a multiplet furnishing a linear representation with four off-shell supersymmetries manifest. 

The calculations are quite cumbersome, but the results are easy to understand: 
By reducing the structure supergroup of eleven-dimensional supergravity, it is possible to separate the torsion constraints of that supergeometry into a part that can be solved off-shell and a remainder that is responsible for putting the theory on-shell. 
Ignoring the latter, and solving the former by the usual method in a top-down approach, gives a collection of so-called {\em reduced field strength superfields}. 
These correspond to our collection of superspace invariants as expressed in \eqref{E:TorsionX} and \eqref{E:TorsionY}. 
These invariants still satisfy a large set of identities expressing, for example, superspace derivatives of one in terms of another. 
Solving these identities results in explicit expressions for the reduced field strengths in terms of the unconstrained superfields \eqref{E:Prepotentials}. 

The miraculous-looking series of fortunate events required in this calculation was guaranteed to have this outcome, because we had previously shown that the set of unconstrained superfields \eqref{E:Prepotentials} correctly describes eleven-dimensional supergravity \cite{Becker:2017zwe}. 
Acting repeatedly with superspace derivatives could, therefore, never generate an equation of motion or an inconsistent constraint, but was guaranteed to construct invariants corresponding to their reduced field strengths, their reduced Bianchi identities, {\it et cetera}.
What was unclear from the outset, was the expression of the torsions in terms of prepotentials, and, in particular, the torsion constraints needed to begin a top-down approach to the problem. 

With the supergeometry torsion constraints so clarified, we could now ``start over'' with the usual superspace Bianchi identity analysis. The advantage of doing this would be a more direct derivation of the super-torsion, super-curvature, and super-4-form. These can be used to define higher-derivative invariants directly that can be used to study deformations of the supergeometry. 

Alternatively, we can proceed to construct higher-derivative deformations of the action directly by making appropriate combinations of the tensors derived herein. (This is related to the approach above, but the methods used in the construction are different.)
Of most direct interest to us are the higher-derivative invariants needed to deform the two-derivative action. 
For M-theory, these are the ones containing $R^4$ terms constructed from the curvature 2-form and those terms related to them by supersymmetry transformations. 
In our hybrid superspace, these are divided into groups related by the manifest supersymmetries which are, in turn, related by the non-manifest ones.
We proceed by constructing $R^4$ invariants of the former that are closed as differential forms (as needed for terms like $\sim \int C \wedge R^4$) in superspace. 

To check that such an approach is feasible, we have carried out the analogous calculation in a five-dimensional version of this scenario in \cite{Becker:2019fri}. 
There the invariant is of the form $\int A \wedge R^2$ where $A$ is the gravi-photon.
(Such an analysis was possible because enough of the 5D hybrid supergeometry had been worked out previously in \cite{Gates:2003qi}.)
A comparison of the 11D and 5D geometries reveals that the latter is embedded in the former. 
This implies that the results of the 5D analysis for $R^2$ can be extended to 11D, as we intend to demonstrate explicitly in future work.

\section*{Acknowledgements}
It is a pleasure to thank 
Jim Gates,
Konstantinos Koutrolikos, 
and Warren Siegel 
for illuminating discussions.
This work is partially supported by National Science Foundation grant PHY-1820921.

\appendix
\section{4D, N=1 Superspace Identities}
\label{S:Superspace}

In this appendix, we rederive the $D$-identities in flat space for Minkowski superspace $\mathbf R^{4|4}$ in the conventions of reference \cite{Buchbinder:1998qv} (which, for most practical calculations, agrees with \cite{Wess:1977fn}).
The bosonic subspace is $\mathbf R^4$ with Minkowski metric (components $\eta_{ab}$ with ${}_{a,b} = {}_{0,1,2,3}$) and a compatible $SL(2, \mathbf C)$ structure (components $\varepsilon_{\alpha \beta}$ and conjugate $\varepsilon_{\dt\alpha \dt \beta}$ with (anti-)Weyl spinor indices ${}_{\alpha, \beta, \dt \alpha, \dt \beta} = {}_{1,2}$). 
These are related by the Pauli matrices $(\sigma_a)_{\alpha \dt \alpha}$ which may be interpreted as the components of isomorphisms $v^a \mapsto v_{\un a} :=v_{\alpha \dt \alpha}:= (\sigma_a)_{\alpha \dt \alpha} v^a$ between vectors and hermitian matrices acting on Weyl spinors. 
The operation is so pervasive in superspace calculations that we use this special notation throughout the paper without qualification. Using it, the compatibility of the Minkowski and $SL(2, \mathbf C)$ structures is expressed by the Fierz identity 
\begin{align}
\label{E:Fierz}
\eta_{\un a \un b} =-2 \varepsilon_{\alpha \beta} \varepsilon_{\dt \alpha \dt \beta}
~~~\Leftrightarrow~~~
(\sigma_a)_{\alpha \dt \alpha} (\sigma_b)_{\beta \dt \beta} \eta^{ab} =-2 \varepsilon_{\alpha \beta} \varepsilon_{\dt \alpha \dt \beta}
\end{align}

\paragraph{$N=1$ algebra} Here we recall some formulae which all follow from the fundamental definition
\begin{align}
\left\{ D_\alpha , \bar D_{\dot \alpha}\right\} = -2i \partial_{\un a}
\end{align}
First we define $\triangle_{\un a}  := \left[ D_\alpha , \bar D_{\dot \alpha}\right]$ and find
\begin{align}
\triangle_{\un a} = \left\{ 
\begin{array}{r}
2i \partial_{\un a} + 2 D_\alpha\bar D_{\dot \alpha}\\
-2i \partial_{\un a} - 2\bar D_{\dot \alpha} D_\alpha
\end{array}
\right.
\end{align}
With three $D$s we have
\begin{subequations}
\begin{align}
[ \bar D^2, D_\alpha]= 4i \partial_{\un a} \bar D^{\dot \alpha} ~~~&,~~~
[ D^2, \bar D_{\dot \alpha}]= -4i \partial_{\un a} D^{\alpha}
\\
\label{E:D*DD*}
\{ \bar D^2, D_\alpha \} = - 2 \bar D_{\dot \alpha}D_{\alpha}\bar D^{\dot \alpha} ~~~&,~~~
\{ D^2, \bar D_{\dot \alpha} \} = - 2 D^{\alpha}\bar D_{\dot \alpha}D_{\alpha}
\\
\label{E:Diamond}
D^\alpha (D_\alpha \bar D_{\dt \alpha} + 2\bar D_{\dt \alpha} D_\alpha ) &= -\bar D_{\dt \alpha} D^2 
\\
(D_\alpha \bar D_{\dt \alpha} + 2\bar D_{\dt \alpha} D_\alpha) \bar D^{\dt \alpha} &= - \bar D^2 D_\alpha
\\
(D_\alpha \bar D_{\dt \alpha} + 2\bar D_{\dt \alpha} D_\alpha) D^\beta \Psi_\beta &= 
	D^2 \bar D_{\dt \alpha} \Psi_\alpha + 2 D^\beta \bar D_{\dt \alpha} D_{(\alpha} \Psi_{\beta)}
\end{align}
\end{subequations}
and usual ones with four
\begin{align}
\left\{ D^2, \bar D^2\right\}=  2D^\gamma\bar D^2 D_\gamma +16 \Box ~~~&,~~~
 \left[ D^2, \bar D^2\right] = -4i \partial^{\un a} \triangle_{\un a}
 \end{align}
 and 
\begin{align}
D_{(\alpha} \bar D^2 D_{\beta)}=  -2i \partial_{(\alpha}{}^{\dt \gamma} \triangle_{\beta) \dt \gamma}
~~~&,~~~
\bar D_{(\dt \alpha} D^2 \bar D_{\dt \beta)}= - 2i \partial^\gamma{}_{(\dt \alpha} \triangle_{\gamma \dt \beta)}
 \end{align}
There are other useful identities involving commutators. First of all, there is the symmetric part of the product
\begin{align}
\label{CommSymm}
\left\{ \triangle_{\un a}, \triangle_{\un b}\right\} &= -8 \partial_{\un a} \partial_{\un b} + 2 \varepsilon_{\alpha \beta} \varepsilon_{\dot \alpha \dot \beta} D^\gamma \bar D^2 D_\gamma 
\end{align}	
and the anti-symmetric part
\begin{align}
\label{CommAnti}
\left[ \triangle_{\un a}, \triangle_{\un b}\right] &= 
	2\varepsilon_{\dot \alpha \dot \beta} D_{(\alpha} \bar D^2 D_{\beta)} 	
	- 2 \varepsilon_{\alpha \beta}\bar D_{(\dot \alpha} D^2 \bar D_{\dot \beta)}
\cr
	&=- 4 i \varepsilon_{\dot \alpha \dot \beta}\partial_{(\alpha}{}^{\dot \gamma}\triangle_{\beta)\dot \gamma}
		+ 4 i \varepsilon_{\alpha \beta}\partial^\gamma{}_{(\dot \alpha}\triangle_{ \gamma \dot \beta)}
\end{align}	
This operator maps $V\mapsto \widetilde F_{ab}(V)$. The imaginary operator
\begin{align}
 -4i \partial_{[{\un a}} \triangle_{{\un b}]} &= \varepsilon_{\dot \alpha \dot \beta} D_{(\alpha} \bar D^2 D_{\beta)} +\varepsilon_{\alpha \beta}\bar D_{(\dot \alpha} D^2 \bar D_{\dot \beta)}\cr
 	&=-2i \varepsilon_{\dot \alpha \dot \beta}\partial_{(\alpha}{}^{\dot \gamma}\triangle_{\beta)\dot \gamma}
		-2i \varepsilon_{\alpha \beta}\partial^\gamma{}_{(\dot \alpha}\triangle_{ \gamma \dot \beta)}
\end{align}	
maps $V\mapsto  F_{ab}(V)$ (up to factors). Finally we point out that the dimension-3 operator
\begin{align}
D^\gamma \bar D^2 D_\gamma \triangle_{\un a} = -8 \partial^{\un b}\partial_{[{\un a}}\triangle_{{\un b}]}
\end{align}
maps $V$ to the component Maxwell equation. In this sense the operator $D^\gamma \bar D^2 D_\gamma$ maps $A_a \mapsto \partial^b F_{ab}(A)$.

\section{\texorpdfstring{$G_2$}{Lie Algebra} Miscellanea}
\label{S:G2}
In this appendix, we review some $G_2$ linear algebra \cite{joyce2000compact, Hitchin:2000jd, Hitchin2001, Bryant2003, Karigiannis:0301218}.

\subsection{Linear Algebra}
\label{S:G2a}
Set $Y = \mathbf R^7$ and let $\varphi \in \Lambda^3(Y)$ be a fixed constant 3-form. 
Define the map $s: \Lambda^3(Y) \to S^2(Y)$ from 3-forms to symmetric tensors by
\begin{align}
\label{E:smap}
s(\omega)_{ij} := -\tfrac1{144} 
	\varepsilon^{k_1\cdots k_7} \varphi_{i k_1k_2} \varphi_{j k_3k_4} \omega_{k_5k_6k_7}
= - \tfrac1{24} \varphi_{i kl} \tilde \omega^{klmn} \varphi_{j mn} 
~.
\end{align}
Then $s(\varphi)_{ij}$ is a symmetric bilinear form.
When this form is invertible, $\varphi$ is said to be {\em stable}.
A stable 3-form on the tangent spaces of $Y$ reduces the structure group $GL(7) \to G_2$ so that $Y$ is a $G_2$-structure manifold.
In this case, $s_{ij}(\varphi)$ is definite, and we may assume it to be positive-definite (by flipping $\varphi\to -\varphi$, if necessary).
When this is true, we may take it to be \cite{Grigorian:2009ge}
\begin{align}
- \varphi= e^{123} + e^{145} 
	+e^{167} + e^{246}
	-e^{257} - e^{347}
	-e^{356}
~.	
\end{align}
Then $s$ is normalized so that $s(\varphi)_{ij} = \delta_{ij}$.

Define the Hodge dual $\psi : = \ast \varphi$. (Recall $\ast^2 = \mathrm {id}$.)
These tensors satisfy the algebraic identities
\begin{subequations}
\label{E:contractions}
\begin{align}
&{\psi}^{ijkl}{\psi}_{i'j'k'l} =
	6\delta^{i}_{[i'} \delta^{j}_{j'}\delta^{k}_{k]}
	-\varphi_{i'j'k'}\varphi^{ijk}
	-9 \delta^{[i}_{[i'}{\psi}_{j'k']}{}^{jk]}
{~},
\\
&
{\psi}^{ijkl} {\psi}_{ijk'l'} = 8 \delta_{[k'}^k \delta_{l']}^l - 2 {\psi}_{k'l'}{}^{kl}
{~},~~
\varphi^{ijk} \varphi_{ij'k'} = 2\delta_{[j'}^j\delta_{k']}^k - {\psi}_{j'k'}{}^{jk}
{~},
\label{E:contractions7projector}
\\
\label{E:psiphi}
&\psi^{ijk l} \varphi_{i'j' l} = 6 \delta^{[i}_{[i'} \varphi_{j']}{}^{jk]} 
{~},~~
\\
\label{E:phiphiclutch}
&\varphi^{ijk} \varphi_{ijk'} = 6\delta_{k'}^k
{~},~~
{\psi}^{ijkl} {\psi}_{ijkl'} = 24 \delta_{l'}^l
{~},~~
\varphi_i{}^{lm} {\psi}_{jk lm} = - 4 \varphi_{ijk}
{~},
\\
\label{E:3phis}
&\epsilon^{lmnpqrs} \varphi_{imn}\varphi_{jpq}\varphi_{krs} = -48 \sqrt{g}\delta^l_{(i} g_{jk)}
~,
\\
&
\varphi_{ijk}\epsilon^{jkl mnpq} = 10 \psi^{[lmnp}\delta^{q]}_i
~,~~\textrm{and}~~~
\psi_{ijkl}\epsilon^{jkl mnpq} = 4! \varphi^{[mnp}\delta^{q]}_i
~.
\end{align}
\end{subequations}
Under the reduction $GL(7)\to G_2$, the $\mathbf{21}$-dimensional space of 2-forms on $Y$ decomposes into $G_2$ representations as $\mathbf{21}=\mathbf{7}\oplus \mathbf{14}$.
Similarly, the $\mathbf{35}$-dimensional space of 3-forms on $Y$ decomposes as $\mathbf{35}=\mathbf{1}\oplus\mathbf{7}\oplus \mathbf{27}$.
For any $p$-form $\omega$, let $\omega_\mathbf i := \pi_{\bm {i}} \omega$ denote the projection to the $\mathbf{i}$-dimensional representation.
Explicitly, for any 2-form $\eta$ and 3-form $\omega$,
\begin{subequations}\label{E:G2Projs}
\begin{align}
\label{E:14to7}
\pi_{\bm 7} \eta_{ij} &= \left( \tfrac13\delta_i^k \delta_j^l  - \tfrac16{\psi}_{ij}{}^{kl} \right) \eta_{kl}
	= \tfrac16 \varphi_{ijk}\varphi^{klm} \eta_{lm}
{~},\\
\label{E:21to14}
\pi_{\bm {14}} \eta_{ij} &= \left( \tfrac23\delta_i^k \delta_j^l  +\tfrac16 {\psi}_{ij}{}^{kl} \right) \eta_{kl}
{~},\\
\label{E:35to1}
\pi_{\bm {1}} \omega_{ijk} &= \tfrac1{42} \varphi_{ijk} \varphi^{i'j'k'} \omega_{i'j'k'}
{~},\\
\label{E:35to7}
\pi_{\bm {7}} \omega_{ijk} &=  \left(
	\tfrac14\delta_i^{i'}\delta_j^{j'}\delta_k^{k'}
	-\tfrac38 {\psi}_{[ij}{}^{i'j'}\delta_{k]}^{k'}
	-\tfrac1{24}\varphi_{ijk} \varphi^{i'j'k'}\right) \omega_{i'j'k'}
	= -\tfrac1{24} \psi_{ijkl}\psi^{lmnp} \omega_{mnp}
{~},\\
\label{E:35to27}
\pi_{\bm {27}} \omega_{ijk} &=  \left(
	\tfrac34\delta_i^{i'}\delta_j^{j'}\delta_k^{k'}
	+\tfrac38 {\psi}_{[ij}{}^{i'j'}\delta_{k]}^{k'}
	+\tfrac1{56}\varphi_{ijk} \varphi^{i'j'k'}\right) \omega_{i'j'k'}
{~~}.	
\end{align}
\end{subequations}
The $\bm 7$-projections of 2- and 3-forms play an important role in the gravitino analysis. We define for such projections the vectors fields\footnote{That is, for any $\eta \in \Lambda^2(Y)$ and $\omega \in \Lambda^3(Y)$, we are defining the vectors $\vec{\eta}$ and $\vec{\omega}$ on $Y$ such that
\begin{align}
\iota_{\vec{\eta}} \varphi = \pi_{\bm 7} \eta
~~~\textrm{and}~~~
\iota_{\vec{\omega}} {\psi} = 2 \pi_{\bm 7} \omega
\end{align}
} 
\begin{subequations}
\begin{align}
\label{E:7components2}
\eta^i:=\tfrac16 \varphi^{ijk} \pi_{\bm 7} \eta_{jk} = \tfrac16 \varphi^{ijk} \eta_{jk}
~~~&\Leftrightarrow~~~ \pi_{\bm 7} \eta_{ij} = \varphi_{ijk} \eta^k
\\
\label{E:7components3}
\omega^i := \tfrac1{12} {\psi}^{ijkl} \pi_{\bm 7} \omega_{jkl}= \tfrac1{12}{\psi}^{ijkl} \omega_{jkl}
~~~&\Leftrightarrow~~~
\pi_{\bm 7} \omega_{ijk} = -\tfrac12 {\psi}_{ijkl} \omega^l
{~~}.
\end{align}
\end{subequations}
Note that this implies that there are conversion factors in squares
\begin{align}
\label{E:squares}
(\eta^i)^2  = \tfrac16  (\pi_{\bm 7} \eta_{ij})^2
~~~\textrm{and}~~~
(\omega^i)^2  = \tfrac16 (\pi_{\bm 7} \omega_{ijk})^2
{~~}.
\end{align}
The dual 4-form ${\psi}$ acts on 2-forms as ${\psi}_{ij}{}^{kl} \eta_{\bm 7 kl} = -4 \eta_{\bm 7 ij}$ and ${\psi}_{ij}{}^{kl} \eta_{\bm {14} kl} = 2\eta_{\bm {14} ij}$, or 
\begin{align}
\label{E:2formSquare}
{\psi}^{ij kl} \eta_{ij} \eta_{kl} 
= -4 \eta_{\bm 7 ij}^2 + 2 \eta_{\bm {14} ij}^2
= -24 (\eta^i)^2 + 2 \eta_{\bm {14} ij}^2
{~~}.
\end{align}

\paragraph{Hitchin metric}
The co-calibration with raised indices $\psi^{ijkl} = \tfrac1{3!} \epsilon^{ijklmnp} \varphi_{mnp}$ acts as a(n indefinite) metric on the space of 2-forms on $Y$. 
There is also a natural (indefinite) metric on the space of 3-forms resulting from the second variation of the Hitchin functional with respect to $\varphi$
\begin{align}
\hyperlabel{E:HitchinMetric}
\hitchin^{ijk}{}_{lmn}&:= \tfrac12 \delta^{[i}_{[l}\delta^j_m\delta^{k]}_{n]} + \tfrac1{36} \varphi^{ijk}\varphi_{lmn} +\tfrac 34 \delta^{[k}_{[l} \psi^{ij]}{}_{mn]}	
~
\end{align}
to which we will refer as the Hitchin metric.\footnote{This definition differs from that of our previous papers by a burdensome factor of 18.}
In terms of projectors \eqref{E:35to1}|\eqref{E:35to27}, it is given by $\hitchin = - \tfrac43 \pi_{\bm {1}} -\pi_{\bm {7}} +\pi_{\bm {27}}$.
Its partial contractions satisfy 
\begin{subequations}
\begin{align}
\varphi_{ijk} \hitchin^{jkl\, mnp} &= - \tfrac43\delta_i^{[l}\varphi^{mnp]}
\\
\psi_{ijkl} \hitchin^{jkl\, mnp} &= - \psi_i{}^{mnp}
\\
\varepsilon_{ijklqrs} \hitchin^{qrs\, mnp} 
&= 
	\tfrac12 \varepsilon_{ijkl}{}^{mnp} 
	+\tfrac16 \psi_{ijkl}\varphi^{mnp}
	+18 \delta_{[i}^{[m}\delta_{j}^{n} \varphi_{kl]}{}^{p]}
\end{align}
\end{subequations}

\paragraph{Torsion of the $G_2$ structure}
As we have reviewed, a stable 3-form $\varphi$ on a smooth 7-manifold $Y$ defines a Riemannian metric $g(\varphi)$ and, therefore, a unique compatible torsion-free connection $\nabla$. 
The exterior derivative of the calibration $\nabla_m \varphi_{ijk}$ need not vanish; without loss of generality, it can be parameterized as
\begin{gather}
\nabla_m \varphi_{ijk} = \psi_{ijkl}T^l_m
~
\end{gather}
in terms of a $7\times 7$ matrix called the {\em torsion of the $G_2$ structure} \cite{Bryant2003}. 

Under the $G_2$ action, this torsion decomposes into $\bm {49}=\bm {1}+\bm {7}+\bm {14}+\bm {27}$.
Since the connection is torsion-free, antisymmetrizing all indices reduces the left-hand side to the exterior derivative of the calibration. 
This is a generic 4-form which may be expanded as 
\begin{subequations}
\label{E:TorsionForms}
\begin{align}
d \varphi = \tau_0 \psi + 3\tau_1 \wedge \varphi + \ast \tau_3
~,
\end{align}
where $\tau_\mu$ is a $\mu$-form with $\mu=0,1,3$ in the $\bm {1}$-, $\bm {7}$-, and $\bm {27}$-dimensional representations, respectively.
(Evidently the $14$-dimensional part drops out.)
Similarly, the differential of the co-calibration can be expanded as
\begin{align}
d\psi = 4 \tau_1 \wedge \psi + \tau_2 \wedge \varphi
~
\end{align}
\end{subequations}
in terms of a 1-form and a 2-form corresponding to $\bm {7}$- and $\bm {14}$-dimensional representations. 
The 1-forms in these expansions are the same because $\psi^{ijkl} \partial_{[m} \psi_{ijkl]}$ is proportional to $\varphi^{ijk}\partial_{[m}\varphi_{ijk]}$. 
It follows that the $49$ components of the $G_2$ torsion $T_i^j$ may be expressed instead in terms of four $\mu$-forms with $\mu=0,1,2,3$.
Explicitly \cite{Bryant2003}
\begin{align}
\label{E:Forms2Torsion}
(\tau_0) &= \tfrac{4}7 T_k{}^k
\cr
(\tau_1)_i &= - \tfrac16 \varphi_{ij}{}^k T_k{}^j
\cr
(\tau_2)_{ij} &= -\tfrac43 T_{[i}{}^k\delta_{j]k} +\tfrac13 \psi_{ijk}{}^l T_l{}^k
\cr
(\tau_3)_{ijk} &= 
	-\tfrac32 (T_{[i}{}^l+T^l{}_{[i})\varphi_{jk]l} 
	+\tfrac37 \varphi_{ijk} T_l{}^l
~.	
\end{align}
Note that $\tau_2$ is in the $\bm {14}$ representation since it is $6\pi_{\bm {14}}$ acting on $T$ considered as a 2-form $T_{[i}{}^k\delta_{j]k}$, and $\tau_3$ is in the $\bm {27}$-dimensional one since the trace of the symmetric part $T_{(i}{}^k\delta_{j)k}$ is projected out.
Conversely, 
\begin{align}
\label{E:Torsion2Forms}
T_i{}^j = \tfrac1{4} \delta_i^j (\tau_0)
	+\varphi_i{}^{jk} (\tau_1)_k 
	-\tfrac1{2} (\tau_2)_i{}^j 
	-\tfrac1{4} \varphi^{jkl} (\tau_3)_{ikl}
~.	
\end{align}

\paragraph{Linearized Riemannian metric}
The map \eqref{E:smap} is $s_{ij}(\omega) = -\frac16 \ast( \varphi_i \wedge \varphi_j \wedge \omega)$, but we can use the identities \eqref{E:contractions} to give alternative forms for it.
For example, when $\omega$ is obtained by expanding a stable 3-form around the calibration, we obtain the symmetrical expression
\begin{align}
\label{E:smap2}
s(\omega)_{ij} = -\tfrac1{3\cdot 144}\varepsilon^{k_1\cdots k_7} 
	\left[
	\omega_{i k_1k_2} \varphi_{j k_3k_4} \varphi_{k_5k_6k_7}
	+\varphi_{i k_1k_2} \omega_{j k_3k_4} \varphi_{k_5k_6k_7}
	+\varphi_{i k_1k_2} \varphi_{j k_3k_4} \omega_{k_5k_6k_7}
	\right]
~.
\end{align}
This is the same map, because the first and third term together are symmetric and evaluate to the second term (with a factor of 2).
Going the other direction, we can use \eqref{E:contractions} to show that either form reduces to the more economical 
\begin{align}
\label{E:smap3}
s(\omega)_{ij} = \tfrac1{6} \varphi_{(i}{}^{kl}\omega_{j)kl}
~.
\end{align}
In this form, it is also clear from \eqref{E:phiphiclutch} and \eqref{E:35to7} that $s$ projects out the $\bm 7$ component of $\omega$.
\begin{align}
\label{E:smapKer}
s \circ \pi_{\bm 7}(\omega) \equiv 0
~. 
\end{align}
This is the entire kernel, as we conclude by counting dimensions. 

Next, define the map $t: \Lambda^1 \otimes \Lambda^\ast_1  \to \Lambda^3$ from vector-valued 1-forms to 3-forms by
\begin{align}
\hyperlabel{E:tmap}
\tmap(M)_{ijk} := \tfrac32 M_{[i}{}^l \varphi_{jk] l}
\end{align}
for any matrix $M \in \Lambda^1 \otimes \Lambda^\ast_1 $.
This matrix space $\Lambda^1 \otimes \Lambda^\ast_1 \simeq \Lambda^2 \oplus S^2$ so its dimension is $\bm{49} = \bm{21} \oplus \bm{28}$ whereas the image is at most $\bm{35}$-dimensional.
This suggests that the $\bm{14}$-dimensional projection of the anti-symmetric part of $M$ does not survive, and it is easily checked that $(\pi_{\bm {14}} \eta)_{[i}{}^l \varphi_{jk] l} \equiv 0$ with the projectors above.
Similarly $(\pi_{\bm {7}} \eta)_{[i}{}^l \varphi_{jk] l} = \eta_{[i}{}^l \varphi_{jk] l}$ as required.
Composing with $s$, we find $s\circ t (M) = - \tfrac13 M_{(i}{}^k\delta_{j)k} - \tfrac16 \delta_{ij} M_k{}^k $, so the symmetric part of $M$ does not come back to itself even up to a coefficient. 
This suggests we define a new map
\begin{align}
\hyperlabel{E:hmap}
h_{ij} := 3 s_{ij} -\tfrac13 \delta_{ij} \delta^{kl} s_{kl} 
~~~\textrm{s.t.}~~~	
h \circ t (M) = M_{(i}{}^k\delta_{j)k}
~.
\end{align}
Note that this map projects out the $\bm 7$ representation by \eqref{E:smapKer}.
The significance of it is that it defines the graviton as a metric fluctuation from the linearized calibration.

\section{Prepotential Transformations and Partial Invariants}
\label{S:Pregauge}
In this section, we collect the linearized transformations of the prepotentials \eqref{E:Prepotentials} found in \cite{Becker:2017zwe}. 
They contain abelian transformations of the M-theory 3-form \cite{Becker:2016xgv}, a non-abelian gauging thereof under 7D diffeomorphisms \cite{Becker:2016rku}, local 4D, $N=1$ superconformal transformations \cite{Gates:1983nr}, and extensions thereof (gravitino transformations) \cite{Becker:2017zwe}.

\paragraph{Abelian Tensor Hierarchy}
Under abelian 2-form symmetry only the prepotentials corresponding to the components of the 3-form transform. Explicitly, \cite{Becker:2016xgv}
\begin{subequations}
\label{E:ATHXf}
\begin{align}
\delta_{ath} \Phi_{ijk} &= 	
	3\partial_{[i}\Lambda_{jk]}
\\
\delta_{ath} V_{ij} &= 
	 \tfrac1{2i} \left( \Lambda_{ij} - \bar \Lambda_{ij} \right) 
	-2 \partial_{[i} u_{j]}
\\
\delta_{ath} \Sigma^\alpha_i &=
	-\tfrac14 \bar D^2 D^\alpha u_i + \partial_i \Upsilon^\alpha 
\\
\delta_{ath} X &=
	 \tfrac1{2i} \left( D^\alpha \Upsilon_\alpha - \bar D_{\dt \alpha} \bar \Upsilon^{\dt \alpha}\right) 
\end{align}
\end{subequations}
where $\Lambda_{ij}$ is chiral, $u_i$ is real, and $\Upsilon^\alpha$ is chiral. Together they define an eleven-dimensional 2-form. 
The field strengths of the abelian tensor hierarchy are 
\begin{subequations}
\begin{align}
\hyperlabel{E:G}
\G &= -\tfrac14 \bar D^2 X 
\\
\hyperlabel{E:H}
\H_i &= \tfrac1{2i} \left( D^\alpha \Sigma_{\alpha i} - \bar D_{\dt \alpha} \bar \Sigma^{\dt \alpha}_i\right) 
	-\partial_i X
\\
\hyperlabel{E:W}
\W_{ij}^\alpha &= -\tfrac14 \bar D^2 D^\alpha V_{ij} 
	+2\partial_{[i} \Sigma_{j]}^\alpha 
\\
\hyperlabel{E:F}
\F_{ijk} &= \tfrac1{2i}\left( \Phi_{ijk} - \bar \Phi_{ijk} \right) 
	-3\partial_{[i} V_{jk]}
\\
\hyperlabel{E:E}
\E_{ijkl} &= ~~4\partial_{[i} \Phi_{jkl]}
\end{align}
\end{subequations}
Note that $\G$, $\W$, and $\E$ are chiral and that $\H$ and $\F$ are real. 
It is almost manifest that these combinations are invariant under the linearized non-abelian gauge transformations \eqref{E:ATHXf}. 

\paragraph{Kaluza-Klein Gauge Field}
The abelian tensor hierarchy was coupled to the non-abelian Kaluza-Klein field gauging the 7D diffeomorphisms in \cite{Becker:2016rku}. 
At the linearized level, this field is described by a real superfield $\mathcal V^i$ carrying a 7D vector index, and transforming as
\begin{align}
\delta_7 \mathcal V^i = \tfrac1{2i} (\tau^i -\bar \tau^i)
~~~:~~~
\bar D_{\dt \alpha} \tau^i =0
~.
\end{align}
Its field strength 
\begin{align}
\hyperlabel{E:KK}
\KK_\alpha^i &= -\tfrac14 \bar D^2 D_\alpha \mathcal V^i 
\end{align}
is invariant and chiral. 

\paragraph{Extended 4D Superconformal Transformations}
The 4D, $N=1$ superconformal transformations are parameterized by a spinor parameter superfield $L^\alpha$ \cite{Gates:1983nr}. 
The prepotential $U^a$ containing the conformal part of the 4D frame suffers the pregauge transformation 
\begin{align}
\delta U^{\alpha \dt \alpha} := \bar D^{\dt \alpha} L^\alpha - D^\alpha \bar L^{\dt \alpha}
\end{align}
where $U^{\un a} = U^{\alpha \dt \alpha} := (\bar \sigma_a)^{\dt \alpha \alpha} U^a$ is by definition the contraction of the vector index by the Pauli matrices (this invertible operation is the same as Feynman's slash but with the diacritical mark directly on the index instead of the field). 
In calculations, it is useful to note that $\bar D L$ appears instead of $L$ itself, so $L^\alpha$ has a gauge-for-gauge ambiguity corresponding to a shift by a chiral spinor field.

As explained in \cite{Linch:2002wg}, this parameter must also enter the transformation of the gravitino superfield:
\begin{align}
\delta \Psi_i^{\alpha} = \Xi_i^\alpha +  D^\alpha \Omega_i + 2i \partial_i L^\alpha
~.
\end{align}
Here $\Xi$ is chiral and $\Omega$ is complex and completely unconstrained. These two parameters are the conformal supersymmetry parameters of the matter gravitino multiplet \cite{Gates:1983nr, Gates:1984mt} (see also appendix C of \cite{Becker:2017zwe} for a review of this multiplet). 

In \cite{Becker:2016edk}, it was observed that the field $X$ in the tensor hierarchy carrying 4D polarizations of the 3-form, was also coupling exactly as the chiral conformal compensator of old-minimal supergravity \cite{Ovrut:1997ur}. This implies that it must transform in a specific way under the $L^\alpha$ part of the superconformal transformations. 
It was also conjectured there that all the fields of the tensor hierarchy suffer similar compensating superconformal transformations.
This is needed for the spectrum of component fields to match as summarized in table \ref{T:Components}. 
This claim was demonstrated explicitly in \cite{Becker:2017zwe} where it was shown that under superconformal transformations, 
\begin{subequations}
\begin{align}
\delta_{sc} \mathcal V^i &= -\tfrac12( \Omega^i + \bar  \Omega^i)
\\
\delta_{sc} X &= D^\alpha L_\alpha + \bar D_{\dt \alpha} \bar L^{\dt \alpha}
\\
\delta_{sc} \Sigma^\alpha_i &= -\Xi^\alpha_i 
\\
\delta_{sc} V_{ij} &= \tfrac1{2i} \varphi_{ijk} (\Omega^k - \bar \Omega^k)
\\
\delta_{sc} \Phi_{ijk} &= -\tfrac i2 \psi_{ijkl} \bar D^2 \bar \Omega^l
~.
\end{align}
\end{subequations}
Crucially, many of these transformations are St\"uckelberg shifts---the hallmark of compensating fields \cite{Gates:1983nr}. 
For the field strengths of the non-abelian tensor hierarchy, this gives
\begin{subequations}
\begin{align}
\delta_{sc} \KK_\alpha^i &= \tfrac18 \bar D^2 D_\alpha ( \Omega^i + \bar  \Omega^i)
\\
\delta_{sc} \G &= -\tfrac14 \bar D^2 D^\alpha L_\alpha
\\
\delta_{sc} \H_{i}&= -\tfrac1{2i} \left( D^\alpha  \Xi_{\alpha i} - \bar D_{\dt \alpha} \bar \Xi_i^{\dt \alpha} \right)
	- \partial_i \left( D^\alpha  L_{\alpha} + \bar D_{\dt \alpha} \bar L^{\dt \alpha} \right)
\\
\delta_{sc} \W^\alpha_{ij} &= 
	\tfrac i8 \varphi_{ijk} \bar D^2 D^\alpha (\Omega^k-\bar\Omega^k) 
	- 2 \partial_{[i} \Xi^\alpha_{j]}
\\
\delta_{sc} \F_{ijk} &= -\tfrac14 \psi_{ijkl} \left( D^2 \Omega^l + \bar D^2 \bar \Omega^l  \right) 
	+\tfrac {3i}2 \varphi_{l[ij} \partial_{k]} (\Omega^l-\bar\Omega^l)
\\
\delta_{sc} \E_{ijkl} &=  - 2i  \psi_{m[ijk} \partial_{l]}\bar D^2 \bar \Omega^m
~.
\end{align}
\end{subequations}

\paragraph{Partial Invariants}
These transformations can be removed by combining them with the superfields $U^a$ and $\Psi_i^\alpha$ requiring the compensation. 
We do this step-wise, first defining partially-invariant building blocks. 
A convenient set is
\begin{subequations}
\begin{align}
\hyperlabel{E:X}
\X_i^{\alpha \dt \alpha} &:= \tfrac 1{2i} ( \bar D^{\dt \alpha}  \Psi_i^\alpha 
	+ D^\alpha \bar \Psi_i^{\dt \alpha}  ) - \partial_i U^{\alpha \dt \alpha}
\\
\hyperlabel{E:T}
\T_i &:= \tfrac1{2i} ( D^\alpha \Psi_{\alpha i} - \bar D_{\dt \alpha} \bar \Psi_i^{\dt \alpha} ) + \H_i
\\
\hyperlabel{E:hatW}
\hatW_{ij}^\alpha &:= \W_{ij}^\alpha +2\partial_{[i} \Psi_{j]}^\alpha
\\
\hyperlabel{E:J}
\J_\alpha^i &:=\KK^i_\alpha - \tfrac i2\varphi^{i jk} \hatW_{\alpha jk}
	+ \tfrac i{12} \psi^{i jkl} \spinorYYY_{\alpha jkl}
\\
\hyperlabel{E:Z}
\Z_{ijk} 
	&:= \F_{ijk} -\tfrac i2 \psi_{ijkl} \T^l 
		- 3i \varphi_{l[ij} \partial_{k]} \mathcal V^l
\end{align}
\end{subequations}
These combinations are invariant under the transformations generated by the $L^\alpha$ and $\Xi_i^\alpha$ parameters:
\begin{subequations}
\begin{align}
\delta \X_{\alpha \dt \alpha}^i &= \tfrac1{2i} \left( \bar D_{\dt \alpha} D_\alpha \Omega^i 
	+ D_\alpha \bar D_{\dt \alpha} \bar\Omega^i 
	\right) 		
\\
\delta \T^{i}&= \tfrac1{2i} \left( D^2 \Omega^i - \bar D^2 \bar \Omega^i \right)
\\
\delta \hatW^\alpha_{ij} &= 
	\tfrac i8 \varphi_{ijk} \bar D^2 D^\alpha (\Omega^k-\bar\Omega^k) 
	+ 2 \partial_{[i} D^\alpha \Omega_{j]}
\\
\delta \J_\alpha^i &= \tfrac14 \bar D^2 D_\alpha \left(  2 \Omega^i -\bar  \Omega^i\right)
\\
\delta \Z_{ijk} &= - \tfrac 12 \psi_{ijkl} D^2 \Omega^l 
	+ 3i \varphi_{l[ij}\partial_{k]}( \Omega^l -i \bar \tau^l)
\end{align}
\end{subequations}
The first of these is a partial covariantization of the vector component of $\Psi$ by the derivative of $U^a$. Alternatively, it is the partial covariantization of the derivative of $U^a$.\footnote{The other two combinations cannot be covariantized under the $L^\alpha$ transformation in this way, because there are no fields analogous to $U^a$ and $X$ transforming into $\bar D \Psi - D\bar \Psi$ and $DL - \bar D\bar L$.} The second is an analog of this for the scalar component of $\Psi$ or the derivative of the compensator $X$. 
$\widehat W$ is just the covariantization of $\W$ with respect to the $\Xi$ parameter.
Next, $\J$ is a combination of the spin-1/2 fields in the $\bm 7$-dimensional representation of $G_2$: the KK-ino, the abelian gaugino, and the tensorino. 
Finally, $\Z$ is the complex combination of scalars appearing in the tensorino $\spinorYYY^\alpha_{ijk}$ \eqref{E:spinorYYY} covariantized by the derivative of the KK prepotential $\mathcal V^i$. Note that it is the only combination transforming under the $\tau^i$ parameter.

Additionally, we define combinations 
\begin{subequations}
\begin{align}
\hyperlabel{E:S}
\S &:=\tfrac12(\G+\bar \G) -\tfrac14 [D_\alpha \bar D_{\dt \alpha} ] U^{\un a}
\\
\hyperlabel{E:P}
\P &:=\tfrac 1{2i} (\G-\bar \G) + \partial_a U^a
\end{align}
\end{subequations}
transforming {\em only} under $L^\alpha$
\begin{subequations}
\begin{align}
\delta \S &
	= \tfrac38 ( D^\alpha \bar D^2 L_\alpha + \bar D_{\dt \alpha} D^2 \bar L^{\dt \alpha} )
	= -\tfrac32 ( D^\alpha \varepsilon_\alpha + \bar D_{\dt \alpha} \bar \varepsilon^{\dt \alpha} )
\\
\label{E:NNMxF}
\delta \P &
	= \tfrac i8 ( D^\alpha \bar D^2 L_\alpha - \bar D_{\dt \alpha} D^2 \bar L^{\dt \alpha} )
	= \tfrac 1{2i} ( D^\alpha \varepsilon_\alpha - \bar D_{\dt \alpha} \bar \varepsilon^{\dt \alpha} )
\end{align}
\end{subequations}
These are essentially a slight rewriting of the 4D, $N=1$ chiral compensator $\G$ that transforms nicely (as new-minimal and virial compensators \cite{Buchbinder:2002gh, Gates:2003cz, Nakayama:2014kua}).

\paragraph{4-form Field Strength}
Taking combinations of superspace derivatives of these partial invariants, we can construct superfields that are invariant at the linearized level under consideration (covariant at non-linear order). 
For example, since the 4D, $N=1$ supersymmetry parameter $\varepsilon^\alpha = -\tfrac14 \bar D^2 L^\alpha$ is chiral, the chiral superfield $\R= -\tfrac 16 \bar D^2 \S$, is a dimension-1 invariant.
In fact, it is the linearized chiral scalar curvature invariant of old-minimal supergravity that contains the complex auxiliary scalar as the leading component in a Taylor expansion and the four-dimensional curvature scalar at higher order \cite{Gates:1983nr,Wess:1992cp}. 

In the modification of old-minimal supergravity employed by eleven-dimensional supergravity, the prepotential of the chiral compensator is real. The effect of this is that one of the real components of the complex auxiliary field becomes (the 4D Hodge dual of) the 4-form curl of a gauge 3-form \cite{Gates:1980ay, Ovrut:1997ur}.
Continuing to make other dimension-1 invariants in this way, we find (among other things) the following combinations
\begin{subequations}
\label{E:4formInvariants}
\begin{align}
\hyperlabel{E:FSXXXX}
\FSXXXX_{abcd}&=3i \epsilon_{abcd} (\R - \bar \R)
\\
\hyperlabel{E:FSXXXY}
\FSXXXY_{abc\, i} &:= \tfrac14 \epsilon_{abc\un d} \left[
		(D^2 +\bar D^2) \X_{i}^{\un d}
		-[D^\delta, \bar D^{\dt \delta}]\T_i 
	 \right]
\\
\tilde \FSXXXY^{\un a}_{i} &:= \tfrac12 [D^\alpha , \bar D^{\dt \alpha}]\T_i - \tfrac12 (D^2 +\bar D^2) \X_{i}^{\un a}
\\
\hyperlabel{E:FSXXYY}
\FSXXYY_{\alpha \beta ij} &:=
	- \tfrac i2 D_{(\alpha} \hatW_{\beta)ij}
	+ \tfrac12 \varphi_{ijk} \partial_{(\beta}{}^{\dt \alpha} \X_{\alpha) \dt \alpha}^k
\\
\hyperlabel{E:FSXYYY}
\FSXYYY_{\un a ijk} 
	&:= \tfrac i2 \left[ \bar D_{\dt \alpha} \spinorYYY_{\alpha ijk}
	+D_\alpha \bar \spinorYYY_{\dt \alpha ijk}\right]
	-3 \varphi_{l [ij} \partial_{k]} \X_{\un a}^l
	\cr
	&= -\tfrac12 \left[ \bar D_{\dt \alpha} D_\alpha \Z_{ijk}
	-D_\alpha \bar D_{\dt \alpha} \bar \Z_{ijk} \right]
	-3 \varphi_{l [ij} \partial_{k]} \X_{\un a}^l
	\cr
	&=\tfrac12 [D_\alpha, \bar D_{\dt \alpha} ] \F_{ijk}
		+ \tfrac12 \psi_{ijkl} \partial_{\un a} \T^l 
		- 3 \varphi_{l[ij} \partial_{k]} ( \X_{\un a}^l - \partial_{\un a} \mathcal V^l)
\\
\hyperlabel{E:FSYYYY}
\FSYYYY_{ijkl}  &:= 2 \partial_{[i} \left[ \Phi_{jkl]} + \bar \Phi_{jkl]} + \psi_{jkl] m}\T^m \right]
~.
\end{align}
\end{subequations}
It can be checked that the new superfields represent the dual of the seven 3-forms $\FSXXXY_{abc i} = \epsilon_{abcd} \tilde \FSXXXY^{d}_{i}$, the 21 2-forms $\FSXXYY_{abij} = (\sigma_{ab})_\alpha{}^\beta \FSXXYY_{ij\beta }{}^\alpha  +$ h.c., the 35 1-forms and the 35 0-form field strengths.

Instead of doing this, we explicitly construct the gauge 3-form in section \ref{S:11DSuperfields} (cf.\ eq.\ \ref{E:gauge3form}). 
Taking the bosonic curl of these components gives an alternative (equivalent) derivation of the dimension-1 4-form invariants.
There, we also define the other eleven-dimensional connections (\ref{E:11Dframe} and \ref{E:grino}) and, through those, the torsion and curvature invariants. 
The off-shell spectrum is completed by auxiliary fields that arise by acting on the connections with supersymmetry transformations. 

\section{Gravitino Descendants} 
\label{S:GrinoDescendants}
The manifest supersymmetry transformations act in superspace by translations in the fermionic directions. 
In this section, we present the result of acting with the fermionic derivatives on the gravitino superfields \eqref{E:grino}.
The process that gives rise to this result is not algorithmic: The desired result is obtained only up to unknown field-dependent gauge transformation and spin connection terms, and in a form that is not easily parsed into 4-form terms and combinations of auxiliary fields. 
We present the solution to this superspace crossword puzzle as irreducible projections of $D$ and $\bar D$ on each of the four gravitino polarizations: The 1-form index may lie along 4D (X) or 7D (Y), and the spinor may have a $G_2$-singlet index (X) or a $\bm 7$-dimensional one (Y).  

The fermionic derivatives of the gravitino with polarization XX may be decomposed as
\begin{subequations}
\begin{align}
D^{(\gamma} \grinoXX_{\un a}{}^{\beta)}  
	&=\omega_{\un a}{}^{\beta \gamma} 
		-\tfrac i2 \delta_\alpha^{(\beta}\varepsilon^{\gamma)\delta} \dX_{\delta\dt \alpha}
\\
D_{\beta}  \grinoXX_{\un a}{}^{\beta} 
	&= \partial_{\un a} N +  i \dX_{\un a} 
\\
\bar D^{\dt \beta}  \grinoXX_{\un a}{}^{\beta} 
	&=  i\delta_\alpha^{ \beta}\delta_{\dt \alpha}^{\dt \beta} \R 
\end{align}
\end{subequations}
The YX descendants are
\begin{subequations}
\begin{align}
D^{(\gamma} \grinoYX_i{}^{\beta)} &= 
	\omega_i{}^{\beta \gamma} 
	- \tfrac i6 \t_i^{\beta \gamma} 
\\
D_{\beta}  \grinoYX_{i}{}^{\beta} 
	&=\partial_i N + i \dY_i  
\\
\bar D^{\dt \beta}  \grinoYX_{i}{}^{\beta } 
	&= -\tfrac i{12} \Ga_i^{\un b} 
\end{align}
\end{subequations}
The XY gravitino is messier:
\begin{subequations}
\begin{align}
D^{(\gamma}  \grinoXY_{\un a}{}^{\beta)j} 
	&= 
	-\tfrac{2i}3 \delta_\alpha^{(\beta}\varepsilon^{\gamma)\delta} \bar \Ga^j_{\delta \dt \alpha}
\\
\label{E:DgrinoXY}
D_{\beta} \grinoXY_{\un a}{}^{\beta j} 
	&=
	\partial_{\un a} N^j 
	+ 2\varphi^j{}_{kl} \omega_{\un a}{}^{kl}
	+\tfrac i2\Ga^j_{\un a}	 	
	+\tfrac{5i}6 \bar \Ga^j_{\un a}
	+\tfrac1{2} \big[
		\bar D_{\dt \alpha} \l^j_{\alpha}
		- D_{\alpha} \bar \l^j_{\dt \alpha }
		\big]
\\
\bar D^{\dt \beta}  \grinoXY_{\un a}{}^{\beta j} 
	&=
	 \partial_{\un a} N^{\un b \,j}
	+ 2 \omega_{\un a}{}^{\un b \,j} 
	+2i \delta_{\dt \alpha}^{\dt \beta} \t^j_{\alpha}{}^{\beta} 
	+\tfrac {2i}3 \delta_\alpha^\beta \bar \t^j_{\dt \alpha}{}^{\dt \beta}
	+2i \delta_\alpha^\beta \delta_{\dt \alpha}^{\dt \beta} \dY^j
\end{align}
\end{subequations}
Finally, the YY part gives
\begin{subequations}
\begin{align}
D^{(\gamma} \grinoYY_{i}{}^{\beta)j} &= 
	- {2i} \pi_{\bm {14}}( \FSXXYY^{\gamma\beta}{}_i{}^j)
\\
D_{\beta} \grinoYY_{i}{}^{\beta j} 
	&= \partial_i N^j 
	+ 2\varphi^j{}_{kl} \, \omega_i{}^{kl}
	+\Ztorsion_i{}^j
\\
\bar D^{\dt \beta} \grinoYY_{i}{}^{\beta j} 
	&= 
	\partial_{i} N^{\un b j}
	+2\omega_i{}^{\un b j}
	- i\gravitonYY_i{}^j (\FSXYYY^{\un b}{}_{klm} )
	+\tfrac 1{6} \varphi_i{}^{jk}\left[
		 \bar D^{\dt \beta} \l_k^\beta
		+i\Ga_k^{\un b} 
	\right]
\end{align}
\end{subequations}
In this derivation, the field-dependent Wess-Zumino gauge transformations are 
\begin{align}
N &:= \tfrac 13 \S-i\P 
\cr
N^i 
&:= \tfrac16 \psi^{ijkl} \left[ \F_{jkl} -\tfrac i2 \psi_{jklm}\T^m\right]
\cr
N^{\un a j} &=
	-\gravitonXY^{\un a j}
	-2i \X^{\un a j} 
	-\bar D^{\dt \alpha}  \Psi^{\alpha j}
	+ D^{\alpha} \bar \Psi^{\dt \alpha j}
~.
\end{align}
We will not need this explicit form, but it is important that the normalizations of the shared parameters agree across equations
(and is an important technical aid in determining the correct decomposition of the descendants). 
The spin connections are given in terms of the frame fields as 
\begin{subequations}
\begin{align}
\omega_{\un c}{}^{\un a\un b}&=\partial^{[\un a} h^{\un b]}_{\un c} 
	-\partial_{\un c} e^{[\un a\un b]}
\\
\hyperlabel{E:spinYXX}
\spinYXX_k{}^{\alpha \beta} &= 
	-\tfrac14 \partial^{(\alpha}{}_{\dt \beta} \gravitonXY_k^{\beta) \dt \beta} 
	-\tfrac18 \partial_k [D^{(\alpha} , \bar D_{\dt \beta} ] U^{\beta)\dt \beta}
=\tfrac14 \varepsilon_{\dt \alpha \dt \beta} \left[ \partial^{[\un a} h_k^{\un b]}
	-\partial_k e^{[\un a\un b]}
	\right]
\\
\omega_{\un c}{}^{\un a \,j} &= \tfrac12 \partial^{\un a} \gravitonXY_{\un c}^j 
	-\tfrac12 \partial^j \gravitonXX_{\un c}^{\un a} 
\\
\omega_i{}^{\un b j} &=\tfrac12 \partial^{\un b} \gravitonYY_i{}^{j} 
	-\tfrac12 \partial^j \gravitonXY_i{}^{\un b} 
\\
\omega_{\un c}{}^{ij} &= \partial^{[i} \gravitonXY_{\un c}^{j]}
\\
\omega_k{}^{ij} &= \partial^{[i} \gravitonYY_k^{j]}
\end{align}
\end{subequations}
These transform as $\delta \omega_{\bm c}{}^{\bm{ab}} = \partial_{\bm c} \lambda^{\bm{ab}}$ with local Lorentz parameters 
\begin{align}
\label{E:localLorentz}
\lambda_{\un a}{}^{\un b} &:=\tfrac 12\left[ 
	\delta_{\dt \alpha}^{\dt \beta} D^{(\beta} \bar D^2 L_{\alpha)} 
	- \delta_\alpha^\beta \bar D^{(\dt \beta} D^2 \bar L_{\dt \alpha)}
\right]
	= -2 \left[ 
	\delta_{\dt \alpha}^{\dt \beta} D^{(\beta} \epsilon_{\alpha)} 
	- \delta_\alpha^\beta \bar D^{(\dt \beta} \bar \epsilon_{\dt \alpha)}
\right]
\cr
\lambda_{\un a}{}^j &:= -\tfrac12 \left[
	\bar D_{\dt \alpha} D_\alpha \Omega^j 
	- D_\alpha \bar D_{\dt \alpha} \bar \Omega^j 
\right]
\cr
\lambda_i{}^{\un b} &:= 
	- \delta_{ij} \varepsilon^{\beta\alpha} \varepsilon^{\dt \beta \dt \alpha} \lambda_{\un a}{}^j
\cr
\lambda_i{}^j &= \tfrac14 \varphi_i{}^{jk} ( D^2 \Omega_k + \bar D^2 \bar \Omega_k) 
	+\tfrac i4 \psi_i{}^{jkl} \partial_k ( \Omega_l - \bar \Omega_l )
\end{align}
In these equations $\Omega^j$ should be interpreted as the quantity $\Omega^j -i \bar \tau$.
This is gauge(-for-gauge)-equivalent to $\Omega^j$ but properly takes into account the non-abelian gauge transformation.


\begin{thebibliography}{10}

\bibitem{Siegel:1981dx}
W.~Siegel and M.~Ro{\v c}ek.
\newblock {On Off-shell Supermultiplets}.
\newblock {\em Phys.Lett.}, B105:275, 1981.
\newblock \href{http://inspirehep.net/record/164156?ln=en}{[{\sc in}SPIRE
  entry]}.

\bibitem{Galperin:2001uw}
A.~S. Galperin, E.~A. Ivanov, V.~I. Ogievetsky, and E.~S. Sokatchev.
\newblock {\em {Harmonic superspace}}.
\newblock Cambridge Monographs on Mathematical Physics. Cambridge University
  Press, 2007.

\bibitem{Galperin:1984av}
A.~Galperin, E.~Ivanov, S.~Kalitsyn, V.~Ogievetsky, and E.~Sokatchev.
\newblock {Unconstrained N=2 Matter, Yang-Mills and Supergravity Theories in
  Harmonic Superspace}.
\newblock {\em Class.Quant.Grav.}, 1:469--498, 1984.
\newblock \href{http://inspirehep.net/record/202528?ln=en}{[{\sc in}SPIRE
  entry]}.

\bibitem{Lindstrom:1989ne}
U.~Lindstr\"om and M.~Ro{\v c}ek.
\newblock {$N=2$ Superyang-mills Theory in Projective Superspace}.
\newblock {\em Commun.Math.Phys.}, 128:191, 1990.
\newblock \href{http://inspirehep.net/record/278888?ln=en}{[{\sc in}SPIRE
  entry]}.

\bibitem{Galperin:1991gk}
A.S. Galperin, Paul~S. Howe, and K.S. Stelle.
\newblock {The Superparticle and the Lorentz group}.
\newblock {\em Nucl. Phys. B}, 368:248--280, 1992.

\bibitem{Delduc:1991ir}
Francois Delduc, Alexander Galperin, and Emery Sokatchev.
\newblock {Lorentz harmonic (super)fields and (super)particles}.
\newblock {\em Nucl. Phys. B}, 368:143--171, 1992.

\bibitem{Berkovits:2000fe}
Nathan Berkovits.
\newblock {Super Poincare covariant quantization of the superstring}.
\newblock {\em JHEP}, 0004:018, 2000.
\newblock \href{http://arXiv.org/abs/hep-th/0001035}{[hep-th/0001035]}.

\bibitem{Cederwall:2010tn}
Martin Cederwall.
\newblock {D=11 supergravity with manifest supersymmetry}.
\newblock {\em Mod. Phys. Lett.}, A25:3201--3212, 2010.
\newblock \href{http://arxiv.org/abs/1001.0112}{[arXiv:1001.0112]}.

\bibitem{Berkovits:2018gbq}
Nathan Berkovits and Max Guillen.
\newblock {Equations of motion from Cederwall\textquoteright{}s pure spinor
  superspace actions}.
\newblock {\em JHEP}, 08:033, 2018.

\bibitem{Siegel:1999ew}
Warren Siegel.
\newblock {Fields}.
\newblock 1999.
\newblock \href{http://arXiv.org/abs/hep-th/9912205}{hep-th/9912205}.

\bibitem{Mandelstam:1982cb}
Stanley Mandelstam.
\newblock {Light Cone Superspace and the Ultraviolet Finiteness of the N=4
  Model}.
\newblock {\em Nucl. Phys.}, B213:149--168, 1983.
\newblock \href{http://inspirehep.net/record/179486?ln=en}{[{\sc in}SPIRE
  entry]}.

\bibitem{Marcus:1983wb}
Neil Marcus, Augusto Sagnotti, and Warren Siegel.
\newblock {Ten-dimensional Supersymmetric {Yang-Mills} Theory in Terms of
  Four-dimensional Superfields}.
\newblock {\em Nucl. Phys.}, B224:159, 1983.
\newblock \href{http://inspirehep.net/record/189742}{[{\sc in}SPIRE entry]}.

\bibitem{Berkovits:1993hx}
Nathan Berkovits.
\newblock {A Ten-dimensional superYang-Mills action with off-shell
  supersymmetry}.
\newblock {\em Phys. Lett.}, B318:104--106, 1993.
\newblock \href{https://arxiv.org/abs/hep-th/9308128}{[arXiv:hep-th/9308128]}.

\bibitem{Evans:1994np}
Jonathan~M. Evans.
\newblock {Supersymmetry algebras and Lorentz invariance for d = 10
  superYang-Mills}.
\newblock {\em Phys. Lett. B}, 334:105--112, 1994.

\bibitem{Becker:2018phr}
Katrin Becker, Melanie Becker, Daniel Butter, and William~D. Linch.
\newblock {$N=1$ supercurrents of eleven-dimensional supergravity}.
\newblock {\em JHEP}, 05:128, 2018.
\newblock \href{https://arxiv.org/abs/1803.00050}{[arXiv:1803.00050]}.

\bibitem{Becker:2017zwe}
Katrin Becker, Melanie Becker, Daniel Butter, Sunny Guha, William~D. Linch, and
  Daniel Robbins.
\newblock {Eleven-Dimensional Supergravity in 4D, $N=1$ Superspace}.
\newblock {\em JHEP}, 11:199, 2017.
\newblock \href{https://arxiv.org/abs/1709.07024}{[arXiv:1709.07024]}.

\bibitem{Becker:2017njd}
Katrin Becker, Melanie Becker, William~D. Linch, III, Stephen Randall, and
  Daniel Robbins.
\newblock {All Chern-Simons Invariants of 4D, N = 1 Gauged Superform
  Hierarchies}.
\newblock {\em JHEP}, 04:103, 2017.
\newblock \href{https://arxiv.org/abs/1702.00799}{[arXiv:1702.00799]}.

\bibitem{Becker:2016edk}
Katrin Becker, Melanie Becker, Sunny Guha, William~D. Linch~III, and Daniel
  Robbins.
\newblock {M-theory potential from the $G_{2}$ Hitchin functional in
  superspace}.
\newblock {\em JHEP}, 12:085, 2016.
\newblock \href{https://arxiv.org/abs/1611.03098}{[arXiv:1611.03098]}.

\bibitem{Becker:2016rku}
Katrin Becker, Melanie Becker, William~D. Linch~III, and Daniel Robbins.
\newblock {Chern-Simons actions and their gaugings in 4D, $N =$ 1 superspace}.
\newblock {\em JHEP}, 06:097, 2016.
\newblock \href{https://arxiv.org/abs/1603.07362}{[arXiv:1603.07362]}.

\bibitem{Becker:2016xgv}
Katrin Becker, Melanie Becker, William~D. Linch~III, and Daniel Robbins.
\newblock {Abelian tensor hierarchy in 4D, $N = 1$ superspace}.
\newblock {\em JHEP}, 03:052, 2016.
\newblock \href{http://arxiv.org/abs/1601.03066}{[arXiv:1601.03066]}.

\bibitem{Gates:1983nr}
S.~James Gates~Jr., Marcus~T. Grisaru, M.~Ro{\v c}ek, and W.~Siegel.
\newblock {\em {Superspace Or One Thousand and One Lessons in Supersymmetry}}.
\newblock Frontiers in Physics, 58. Benjamin/Cummings, 1983.
\newblock \href{http://arXiv.org/abs/hep-th/0108200}{[hep-th/0108200]}.

\bibitem{Wess:1992cp}
J.~Wess and J.~Bagger.
\newblock {\em {Supersymmetry and supergravity}}.
\newblock Princeton, USA: Univ. Pr. (1992) 259 p, 1992.
\newblock \href{http://inspirehep.net/record/350988?ln=en}{[{\sc in}SPIRE
  entry]}.

\bibitem{Buchbinder:1998qv}
I.L. Buchbinder and S.M. Kuzenko.
\newblock {\em {Ideas and methods of supersymmetry and supergravity: Or a walk
  through superspace}}.
\newblock Bristol, UK: IOP (1998) 656 p, 1998.
\newblock \href{http://inspirehep.net/record/485478?ln=en}{[{\sc in}SPIRE
  entry]}.

  
\bibitem{Howe:1997he}
Paul~S. Howe.
\newblock {Weyl superspace}.
\newblock {\em Phys. Lett. B}, 415:149--155, 1997.
\newblock \href{https://inspirehep.net/literature/446234}{[{\sc in}SPIRE entry]}.


\bibitem{Cederwall:2000ye}
Martin Cederwall, Ulf Gran, Mikkel Nielsen, and Bengt~E.W. Nilsson.
\newblock {Manifestly supersymmetric M-theory}.
\newblock {\em JHEP}, 0010:041, 2000.
\newblock \href{https://inspirehep.net/literature/529792}{[{\sc in}SPIRE entry]}.


\bibitem{Nishino:1996tw}
Hitoshi Nishino and S.~James {Gates, Jr.}
\newblock {Toward an off-shell 11D supergravity limit of M-theory}.
\newblock {\em Phys. Lett. B}, 388:504--511, 1996.
\newblock \href{https://inspirehep.net/literature/415718}{[{\sc in}SPIRE entry]}.



\bibitem{Gates:2001hf}
S.~James {Gates, Jr.} and Hitoshi Nishino.
\newblock {Deliberations on 11D superspace for the M-theory effective action}.
\newblock {\em Phys. Lett. B}, 508:155--167, 2001.
\newblock \href{https://inspirehep.net/literature/551915}{[{\sc in}SPIRE entry]}.


\bibitem{Gates:2001zz}
S.~James {Gates, Jr.}
\newblock {Superconformal symmetry in 11D superspace and the M-theory effective action}.
\newblock {\em Nucl. Phys. B}, 616:85--105, 2001.
\newblock \href{https://inspirehep.net/literature/558607}{[{\sc in}SPIRE entry]}.



\bibitem{Becker:2020hym}
Katrin Becker and Daniel Butter.
\newblock {4D $N=1$ Kaluza-Klein superspace}.
\newblock {\em JHEP}, 09:091, 2020.
\newblock \href{http://arXiv.org/abs/2003.01790}{[arXiv:2003.01790]}.

\bibitem{Cremmer:1978ds}
E.~Cremmer and B.~Julia.
\newblock {The N=8 Supergravity Theory. 1. The Lagrangian}.
\newblock {\em Phys.Lett.}, B80:48, 1978.
\newblock \href{http://inspirehep.net/record/131572?ln=en}{[{\sc in}SPIRE
  entry]}.

\bibitem{Gates:1980ay}
S.~James Gates~Jr.
\newblock {Super $p$-Form Gauge Superfields}.
\newblock {\em Nucl.Phys.}, B184:381, 1981.
\newblock \href{http://inspirehep.net/record/9990?ln=en}{[{\sc in}SPIRE
  entry]}.

\bibitem{Ovrut:1997ur}
Burt~A. Ovrut and Daniel Waldram.
\newblock {Membranes and three form supergravity}.
\newblock {\em Nucl. Phys.}, B506:236--266, 1997.
\newblock \href{https://arxiv.org/abs/hep-th/9704045}{[arXiv:hep-th/9704045]}.

\bibitem{Lott:1990zh}
J.~Lott.
\newblock {Torsion constraints in supergeometry}.
\newblock {\em Commun.Math.Phys.}, 133:563--615, 1990.
\newblock \href{http://inspirehep.net/record/305705?ln=en}{[{\sc in}SPIRE
  entry]}.

\bibitem{Lott:2001st}
John Lott.
\newblock {The Geometry of supergravity torsion constraints}.
\newblock 2001.
\newblock \href{http://arxiv.org/abs/math/0108125}{[arXiv:math/0108125]}.

\bibitem{book:976236}
Lorenz~J. Schwachh{\"o}fer.
\newblock {\em Global Differential Geometry}.
\newblock Springer Proceedings in Mathematics 17. Springer-Verlag Berlin
  Heidelberg, 1 edition, 2012.

\bibitem{Figueroa-OFarrill:2015rfh}
Jos{\'e} Figueroa-O'Farrill and Andrea Santi.
\newblock {Spencer cohomology and 11-dimensional supergravity}.
\newblock {\em Commun. Math. Phys.}, 349(2):627--660, 2017.

\bibitem{Becker:2019fri}
Katrin Becker, Melanie Becker, Daniel Butter, William~D. Linch, and Stephen
  Randall.
\newblock {Five-dimensional Supergravity in $N = 1/2$ Superspace}.
\newblock {\em JHEP}, 03:098, 2020.
\newblock \href{https://arxiv.org/abs/1909.09208}{[arXiv:1909.09208]}.

\bibitem{Gates:2003qi}
S.~James Gates~Jr., William D.~Linch III, and J.~Phillips.
\newblock {Field strengths of linearized 5D, $\mathcal N=1$ superfield
  supergravity on a 3-brane}.
\newblock {\em JHEP}, 0502:036, 2005.
\newblock \href{http://arxiv.org/abs/hep-th/0311153}{[hep-th/0311153]}.

\bibitem{Wess:1977fn}
J.~Wess and B.~Zumino.
\newblock {Superspace Formulation of Supergravity}.
\newblock {\em Phys. Lett.}, B66:361--364, 1977.
\newblock \href{http://inspirehep.net/record/123994?ln=en}{[{\sc in}SPIRE
  entry]}.

\bibitem{joyce2000compact}
D.D. Joyce.
\newblock {\em Compact Manifolds with Special Holonomy}.
\newblock Oxford mathematical monographs. Oxford University Press, 2000.

\bibitem{Hitchin:2000jd}
Nigel~J. Hitchin.
\newblock {The Geometry of Three-Forms in Six and Seven Dimensions}.
\newblock {\em J. Diff. Geom.}, 55(3):547--576, 2000.
\newblock \href{http://arxiv.org/abs/math/0010054}{[arXiv:math/0010054]}.

\bibitem{Hitchin2001}
Nigel~J. Hitchin.
\newblock {Stable Forms and Special Metrics}.
\newblock 2001.
\newblock \href{http://arxiv.org/abs/math/0107101}{[arXiv:math/0107101]}.

\bibitem{Bryant2003}
R.~L. {Bryant}.
\newblock {Some remarks on $G_2$-structures}.
\newblock {\em ArXiv Mathematics e-prints}, May 2003.
\newblock \href{https://arxiv.org/abs/math/0305124}{[arXiv:math/0305124]}.

\bibitem{Karigiannis:0301218}
Spiro Karigiannis.
\newblock {\em Deformations of ${G}_2$ and ${Spin}(7)$ Structures on
  Manifolds}.
\newblock PhD thesis, 2003.
\newblock \href{https://arxiv.org/abs/math/0301218}{[arXiv:math/0301218]}.

\bibitem{Grigorian:2009ge}
Sergey Grigorian.
\newblock {Moduli spaces of G(2) manifolds}.
\newblock {\em Rev. Math. Phys.}, 22:1061--1097, 2010.
\newblock \href{https://arxiv.org/abs/0911.2185}{[arXiv:0911.2185]}.

\bibitem{Linch:2002wg}
William D.~Linch III, Markus~A. Luty, and J.~Phillips.
\newblock {Five-dimensional supergravity in $\mathcal N=1$ superspace}.
\newblock {\em Phys.Rev.}, D68:025008, 2003.
\newblock \href{http://arxiv.org/abs/hep-th/0209060}{[hep-th/0209060]}.

\bibitem{Gates:1984mt}
S.~James Gates~Jr. and V.~Alan Kostelecky.
\newblock {Supersymmetric Matter Gravitino Multiplets}.
\newblock {\em Nucl. Phys.}, B248:570--588, 1984.
\newblock \href{http://inspirehep.net/record/15327}{[{\sc in}SPIRE entry]}.

\bibitem{Buchbinder:2002gh}
I.~L. Buchbinder, S.~James Gates~Jr., William~D. Linch~III, and J.~Phillips.
\newblock {New 4D, $N=1$ superfield theory: Model of free massive superspin-3/2
  multiplet}.
\newblock {\em Phys. Lett.}, B535:280--288, 2002.
\newblock \href{http://arxiv.org/abs/hep-th/0201096}{[arXiv:hep-th/0201096]}.

\bibitem{Gates:2003cz}
S.~James Gates~Jr., Sergei~M. Kuzenko, and J.~Phillips.
\newblock {The Off-shell (3/2, 2) supermultiplets revisited}.
\newblock {\em Phys.Lett.}, B576:97--106, 2003.
\newblock \href{https://arxiv.org/abs/hep-th/0306288}{[arXiv:hep-th/0306288]}.

\bibitem{Nakayama:2014kua}
Yu~Nakayama.
\newblock {Imaginary supergravity or Virial supergravity?}
\newblock {\em Nucl. Phys.}, B892:288--305, 2015.
\newblock \href{https://arxiv.org/abs/1411.1057}{[arXiv:1411.1057]}.





\end{thebibliography}

\end{document}